\documentclass[%
 twocolumn,
 superscriptaddress,
 longbibliography,
 nofootinbib,
 amsmath,amssymb,
 aps,
prb,
]{revtex4-2}

\usepackage{graphicx}
\usepackage{dcolumn}
\usepackage{bm}

\usepackage{float}
\usepackage{array}
\usepackage{url}
\usepackage{xcolor}
\usepackage{color}
\usepackage{graphicx}
\usepackage{amsfonts}
\usepackage{amsmath,amssymb}
\usepackage{wasysym}
\usepackage{bbold}
\usepackage{physics}
\usepackage{caption}
\usepackage{subcaption}
\usepackage[normalem]{ulem}
\usepackage{hyperref}
\hypersetup{colorlinks=true,citecolor=blue,linkcolor=blue, urlcolor=blue, breaklinks=true}
\usepackage{mhchem}
\usepackage{mathtools}
\usepackage{multirow}

\newcommand{\chiof}[3]{\ensuremath{(\hat{\va{S}}_{#1}\cross \hat{\va{S}}_{#2})\vdot \hat{\va{S}}_{#3}}}
\newcommand{\Sof}[1]{\ensuremath{\hat{\va{S}}_{#1}}}

\newcommand{\dotof}[2]{\ensuremath{\hat{\va{S}}_{#1}\vdot \hat{\va{S}}_{#2}}}

\newcommand{\upnchi}{\ensuremath{\hat{\chi}_{\triangle,\vec{n}}}}
\newcommand{\downnchi}{\ensuremath{\hat{\chi}_{\triangledown,\vec{n}}}}

\newcommand{\mysymb}[2]{\mathord{\vcenter{\hbox{\includegraphics[height=#2 ex]{#1}}}}}
\newcommand{\Z}{\mathbb{Z}}
\newcommand{\Tm}{\hat{\mathcal{T}}_R}
\newcommand{\tnew}{\mathrm{t}}
\newcommand{\tot}[1]{#1_{\mathrm{tot}}}

\begin{document}

\preprint{APS/123-QED}

\title{Monopole Josephson Effects in a Dirac Spin Liquid}
\author{Gautam Nambiar}
\affiliation{Joint Quantum Institute, Department of Physics, University of Maryland,
College Park, Maryland 20742, USA}
\author{Daniel Bulmash}
\affiliation{Joint Quantum Institute, Department of Physics, University of Maryland,
College Park, Maryland 20742, USA}
\affiliation{Condensed Matter Theory Center, Department of Physics, University of Maryland, College Park, MD 20742, USA}
\author{Victor Galitski}
\affiliation{Joint Quantum Institute, Department of Physics, University of Maryland,
College Park, Maryland 20742, USA}
\date{\today}

\begin{abstract}
Dirac Spin liquids (DSLs) are gapless featureless states, yet interesting by virtue of the effective field theory describing them - (2+1)-dimensional quantum electrodynamics (QED$_3$). Further, a DSL is known to be a ``parent state'' of various seemingly unrelated  ordered states, such as antiferromagnets and valence bond solids in the sense that one can obtain ordered states by condensing magnetic monopoles of the emergent gauge field. Can operators in the effective field theory, such as the emergent electric field, be externally induced and measured? In this work, we exploit the parent state picture to argue that the answer is yes. We propose a range of ``monopole Josephson effects'' that arise when two ordered states are separated by a region of the parent DSL. In particular, we show that one can induce an AC monopole Josephson effect, which manifests  itself as an AC emergent  electric field in the spin liquid, accompanied by a measurable spin current. Further, we show that this AC emergent electric field can be measured as a sharp tunable peak in Raman scattering. This work provides a theoretical proof of principle that emergent gauge fields in spin liquids can be externally induced, manipulated, and probed using more conventional states, which offers a generic platform for studying the exotic spin phases.
\end{abstract}

\maketitle


\section{Introduction}
\label{sec:intro}
Consider a spin-$1/2$ system in its ground state. Flipping a single spin creates a spin-1 excitation. If the ground state is conventional,  such an excitation would disperse creating a superposition of spin-wave modes with spin 1. However, there is strong theoretical reason~\cite{hastings2004lieb,lieb1961two} to expect exotic systems where, in addition to creating spin-1 modes, the spin-flip can fractionalize into two spin-1/2 excitations, which can then move away from each other. One interesting class of such systems in 2+1D are Dirac spin liquids (DSLs).
The effective field theory describing DSLs is usually written in terms of Dirac fermions strongly coupled to an emergent $U(1)$ gauge field. This strongly coupled theory is believed to flow at low energies to a conformally invariant fixed point QED$_3$~\cite{rantner2001electron,hermele2004stability,karthik2016scale}. 

To detect such an exotic state in a given physical system, say a material with spins, we would need to probe the low energy degrees of freedom of the effective field theory describing the state in question.
For example, the low energy excitations of gapped spin liquids in 2+1D are anyonic quasiparticles. There have been proposals in the past for accessing emergent degrees of freedom in such gapped spin liquids with the assistance of more conventional ordered phases, which is helpful because one typically has better control over ordered phases. Examples of these include Refs.~\cite{senthil2001prl,senthil2001prb,barkeshli2014coherent} for $\Z_2$ spin liquids, and Ref.~\cite{aasen2020electrical} for Kitaev spin liquids.

For a strongly coupled field theory in 2+1D, such as QED$_3$ on the other hand, the low energy degrees of freedom are not well-defined quasiparticles, but instead the primary operators of the conformal field theory (CFT). Previous works have proposed ways to measure correlation functions of such operators in the ground state of a DSL~\cite{rantner2001electron,ko2010raman,lee2013proposal}. However there appears to be a lack of proposals to directly control such operators externally and measure them. In this paper, we explore this direction and propose a way to induce and measure an emergent electric field in a DSL. Our proposal relies crucially on coupling to monopole operators.

Monopole operators insert an integer multiple of $2\pi$ flux of the emergent $U(1)$ gauge field. Polyakov showed ~\cite{polyakov1977quark} that in 2+1 D, monopoles are always relevant in the renormalization group (RG) sense and proliferate in a pure $U(1)$ gauge theory, leading to confinement of test charges. Such a theory would not describe a spin liquid phase. However, including gapless fermions in the theory increases the scaling dimension of the monopoles, and in the limit of  large number of fermion flavors $N_f$, monopoles can become irrelevant~\cite{borokhov2003topological,hermele2004stability}. Indeed, using a symmetry analysis followed by a large $N_f$ analysis, Ref.~\cite{song2020spinon} found that on the triangular lattice, monopoles are either disallowed by symmetry or irrelevant, suggesting that a DSL could be a stable phase. Such a phase has an emergent $U(1)_{\text{top}}$ symmetry corresponding to the conservation of total emergent flux (the subscript ``top'' is used to differentiate $U(1)_{\text{top}}$ from $U(1)$ gauge redundancy). In fact, monopoles are charged under an enlarged emergent internal symmetry $G_{IR}=SO(6)\times U(1)_{\text{top}}/\mathbb{Z}_2$ (see Sec.~\ref{sec:overview} for a review). 
 Because spatial symmetries have a nontrivial action in $G_{IR}$, the monopoles transform under the microscopic symmetries like order parameters for magnetic orders including the $120^\circ$ antiferromagnet and the $\sqrt{12}\times \sqrt{12}$ valence bond solid. Therefore, if the $2\pi$ monopoles somehow do proliferate (say as a result of spontaneous symmetry breaking, or due to a symmetry-breaking perturbation), the system exits the DSL phase. The resulting phase is an ordered phase determined by which combination of monopoles proliferates. In this sense, it was suggested that the DSL is a parent state for several seemingly unrelated magnetic and VBS orders~\cite{hermele2005algebraic,song2019unifying}. 

\begin{figure}[h!]
    \centering
    \includegraphics[width=0.45\textwidth]{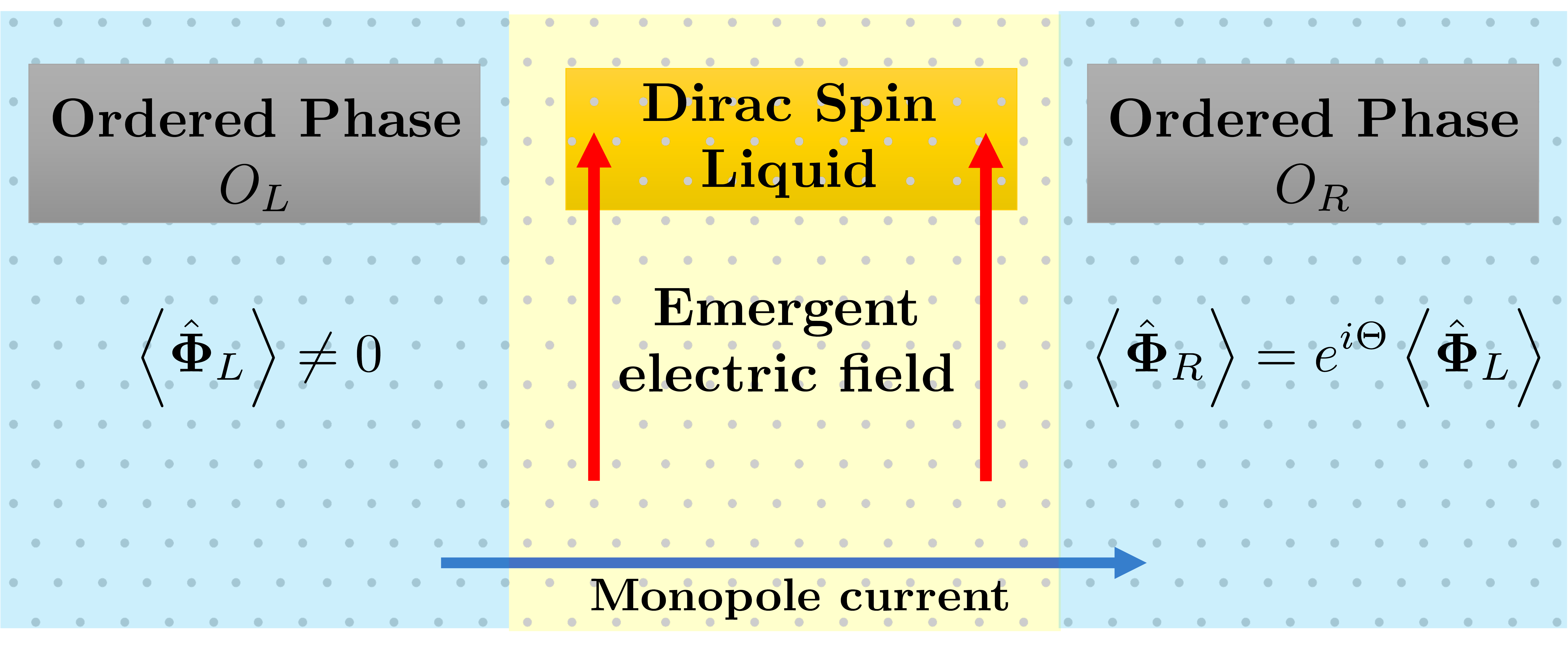}
    \caption{Monopole Josephson effect, general idea: We consider a junction of two ordered phases ($O_L$ and $O_R$) separated by a DSL. $O_L$ and $O_R$ are viewed as monopole condensates such that their expectation values are related by a generalized phase (unitary matrix) $e^{i\Theta}$ (possibly time-dependent). This leads to a monopole current across the junction, which is equivalent to an electric field in the emergent $U(1)$ gauge field in the perpendicular direction inside the DSL.}
    \label{fig:schematic1}
\end{figure}

Can this parent state picture guide us towards finding experimental probes for the low energy theory (QED$_3$) that describes the DSL? Can ordered states in proximity to a spin liquid have an interesting effect on the spin liquid, and vice-versa? In this paper, we argue that the answer to both these questions is yes, by proposing a Josephson junction-like setup shown in Fig.~\ref{fig:schematic1} with two ordered phases separated by a middle region in the DSL phase. The main idea is that since the ordered states can be viewed as monopole condensates, monopoles can tunnel between the ordered states through the DSL. 

We show that in certain circumstances, applying a Zeeman field gradient across the junction has the same effect as a voltage difference across a regular Josephson junction (between superconductors) and thus gives rise to an AC monopole current flowing across the DSL. In $2+1$ dimensions, a monopole current is equivalent to an electric field but in the perpendicular direction. Therefore, this ``monopole Josephson effect'' provides a way to externally induce an emergent electric field through the DSL. We suggest a way to measure the AC emergent electric field optically as a peak in Raman scattering intensity by identifying microscopic operators corresponding to the emergent electric field. In addition to this signature within the DSL region, we show that when the ordered phases are $120^\circ$ antiferromagnets, the same monopole Josephson effect leads to a spin current across the junction. This spin current can in principle be measured on both the ordered side and the DSL using techniques proposed in~\cite{chatterjee2015probing,chen2013detection}. We also discuss three other conceptually related effects which all fall under the umbrella of the monopole Josephson effect.

The rest of the paper is organized as follows. In Sec.~\ref{sec:overview}, we provide a brief review of DSLs, emphasizing the relevant features of monopole operators and the parent state picture. In Sec.~\ref{sec:josephson}, we show that an emergent electric field in the DSL can be induced via the monopole Josephson effect. In Sec.~\ref{sec:raman}, we propose a way to detect this emergent electric field using Raman scattering. In Sec.~\ref{sec:othereff}, we discuss other phenomena related to the monopole Josephson effect. Finally, we offer some general conclusions and discussion in Sec.~\ref{sec:discussion}. 

\section{Review of Dirac Spin Liquids}
\label{sec:overview}
DSLs are described by an effective field theory with $N_f=4$ flavors of gapless Dirac fermions at zero equilibrium density strongly coupled to a compact $U(1)$ gauge field. The fermions carry spin-$1/2$ under the microscopic $SO(3)$ spin rotation symmetry. One way to get to this theory is the parton construction, which we will briefly review below.

Before proceeding, we explain some notation. We use $\vec{V}$ for 2-component vectors in real-space, $\va{V}$ for 3-component vectors in internal spin or valley space, and $\vb{V}$ for vectors in other internal spaces such as $SO(6)$ (see Appendix~\ref{app:notation} for other remarks on notation).

Consider a spin system whose microscopic Hilbert space consists of spin-$1/2$ at each lattice site. We assume that the Hamiltonian realizes a DSL ground state and respects some set of symmetries $G_{UV}$ which include lattice symmetries, time-reversal, and spin-rotation symmetry. We will work on the triangular lattice for concreteness, but the results are general except where otherwise noted. The objective of the parton construction is to come up with a mean-field theory even in the case when the spin operators $\expval{\Sof{i}}=0$ (likewise for other local spin operators) for all sites $i$. In this approach, the Hilbert space at each site $i$ is doubled by writing spin operators in terms of fictitious spin-$1/2$ fermionic  ``spinon'' operators $\hat{f}_{i,\alpha}$: 
\begin{equation}
\label{eq:parton101}
    \Sof{i}=\frac{1}{2} \sum_{\alpha, \beta \in \{\uparrow,\downarrow\}} \hat{f}_{i\alpha}^{\dagger}\vec{\boldsymbol\sigma}_{\alpha \beta} \hat{f}_{i\beta}.
\end{equation}
Here $\vec{\sigma}$ is a vector of Pauli matrices. This description has a $U(1)$ gauge redundancy 
\begin{equation}
    \hat{f}_{i\alpha}\to e^{i\lambda_i} \hat{f}_{i\alpha},
\end{equation}
in the sense that the physical spin operators are invariant under such a transformation. The spin Hilbert space is recovered by imposing the constraint that there is exactly one fermion at each site. One now rewrites the spin Lagrangian in terms of these fermions. A quadratic term in spins becomes a quartic term in fermions, which is then decoupled so that the fermion hopping coefficients $\expval{\sum_{\alpha=\uparrow,\downarrow}\hat{f}_{i\alpha}^\dagger \hat{f}_{j\alpha}}$ acquire mean-field expectation value $\chi_{ij}$. For a DSL on the triangular lattice, the mean-field configuration for $\{\chi_{ij}\}$ consists of alternating $\pi$-flux and $0$-flux on upward and downward triangles respectively. Diagonalizing this quadratic mean-field Hamiltonian gives a spectrum with two Dirac cones (valleys). To zeroth order, the single-occupancy-per-site constraint is relaxed to demanding single-occupancy on average, i.e., that the fermions are at half filling. This forces the chemical potential to lie exactly at the Dirac points. So, to zeroth order, the low-energy theory has 4 flavors of Dirac fermions --- 2 valleys for each spin. $U(1)$ gauge fluctuations $\chi_{ij}\to \chi_{ij}e^{i a_{ij}}$ are now reintroduced. The single particle per site constraint is reintroduced only weakly as a Gauss's law by assuming a finite coupling constant $g$: 
\begin{equation}
\label{eq:gauss}
    \sum_{\alpha=\uparrow,\downarrow}\hat{f}_{i\alpha}^\dagger \hat{f}_{i\alpha}-1 = \frac{1}{g^2} \sum_j \hat{e}_{ij},
\end{equation}
where $e_{ij}$ is the emergent electric field, i.e., the electric flux of $a_{ij}$. This leads us to a field theory Lagrangian density, which schematically is
\begin{equation}
\label{eq:lqed}
    \mathcal{L}=\sum_{i=1}^{N_f=4} \bar{\psi}_i \gamma^\mu \left(\partial_\mu - i a_\mu\right)\psi_i + \frac{1}{8\pi g^2}(\epsilon^{\mu \nu \lambda} \partial_\mu a_{\nu})^2,
\end{equation}
where each $\psi_i$ is a 2-component spinor and $a_0$ has been introduced as a Lagrange multiplier enforcing Eq.~\ref{eq:gauss}. The theory of the DSL is then described by the low-energy fixed point, called QED$_3$, of Eq.~\eqref{eq:lqed}. By itself, Eq.~\eqref{eq:lqed} is not useful to understand the fixed point because the coupling constant $g^2$ has dimensions of $\left[\text{Length}^{-1}\right]$. So, at low energies $g^2$ flows to $\infty$ making gauge fluctuations uncontrolled. However, progress can be made by treating $1/N_f$ as a small parameter. Then, because of screening of gauge fluctuations by the many gapless fermions, $g^2$ approaches a fixed point value which scales as $\Lambda/N_f$ where $\Lambda$ is an inverse length scale of order of the lattice spacing~\cite{hermele2004stability}. Most of the current understanding of the QED$_3$ fixed point comes from this limit --- thinking of the spinons $\psi_i$ essentially as almost free fermions with gauge fluctuations controlled by the large $N_f$ expansion. At the same time, one should keep in mind that the most important low energy operators to study are primary operators of the CFT with the lowest scaling dimensions. 

The essential features of the nontrivial fixed point theory QED$_3$ are: 

\begin{enumerate}
    \item \underline{Monopole operators:} Among the primary operators in QED$_3$ are magnetic monopoles $\hat{\Phi}^\dagger_i$. These operators insert $2\pi$ flux of the emergent gauge field $a$, that is,
    \begin{equation}
    \comm{\hat{b}(x)}{\hat{\Phi}^\dagger_i(x^\prime)}=2\pi \delta(x-x^\prime) \hat{\Phi}^{\dagger}_i (x^\prime),
    \end{equation}
    where $\hat{b}(x)=(\partial_1 \hat{a}_2-\partial_2 \hat{a}_1$)(x). In a path integral, the insertion of these operators corresponds to instanton events whose role is to restore the compactness of the gauge field in the low energy theory. 
    \item \underline{Enlarged emergent symmetry group:}~\cite{song2019unifying,song2020spinon} While the microscopic Hamiltonian has the symmetries listed above, the DSL theory (QED$_3$) has an enlarged internal symmetry $G_{IR}=\frac{SO(6)\times U(1)_{\text{top}}}{\mathbb{Z}_2}$. The $U(1)_{\text{top}}$ symmetry corresponds to the conservation of total emergent magnetic flux through the plane:  $\tot{\hat{b}} \equiv\frac{1}{2\pi}\int d^2x \hat{b}(x)$. Clearly monopole operators are charged under $U(1)_{\text{top}}$. The total flux is conserved because the monopole operators (i.e. flux creation/annihilation operators) have zero expectation value in the wave function described by DSL theory. The $SO(6)$ symmetry corresponds to the internal rotation between the spin and valley indices, and the monopole operators transform as a vector under $SO(6)$.
    
    More concretely, inserting $2\pi$ flux leads to one Landau zero-mode per fermion flavor, and to maintain half filling, 2 out of 4 zero-modes need to be filled. The resulting 6 choices lead to 6 independent monopole creation operators $\hat{\Phi}_i^\dagger$ ($i\in \{1,\ldots 6\}$) which together transform as a vector under $SO(6)$. A complementary way to understand this is to observe that the fermionic partons enjoy an $SU(4)$ symmetry near the Dirac points. Upon carefully keeping track of redundant factors of $\Z_2$\footnote{The $\Z_2$ subgroup of $SU(4)$ generated by fermion parity is actually a $U(1)$ gauge transformation rather than a symmetry, reducing the $SU(4)$ symmetry to $SO(6) \cong SU(4)/\Z_2$. The element $-1 \in SO(6)$ is identical to a $\pi$ rotation in $U(1)_{top}$.}, one arrives at $G_{IR}$ above. While the microscopic $SO(3)\subset G_{UV}$ spin-rotation symmetry is directly $SO(3)_{\text{spin}}\subset G_{IR}$, elements of the space group generally embed nontrivially into $SO(3)_{\text{valley}} \times U(1)_{\text{top}} \subset G_{IR}$ (in addition to the spatial transformation). 
    
    \item \underline{Parent state of competing orders}~\cite{hermele2005algebraic,song2019unifying,song2020spinon} 
    If a monopole operator condenses, i.e., acquires a nonzero expectation value, then the spinons confine, and the low-energy excitations are un-fractionalized spin-1 modes. The resultant phase is simply a conventional magnetically ordered phase. Many seemingly-unrelated ordered phases can appear depending on which monopole operator condenses and what the microscopic symmetries are. For example, on the triangular lattice, a $120^{\circ}$ coplanar order can be obtained by condensing spin triplet monopoles, and valence bond solids with a unit-cell area of 12 times the elementary unit cell can be obtained by condensing spin singlet monopoles. The DSL thus serves as a ``parent state'' for many ordered states, in the sense that one mechanism (monopole condensation) in the DSL is responsible for driving transitions to many different ordered states.
    
    Such a transition could happen for multiple reasons. A monopole operator may be relevant in the renormalization group (RG) sense and symmetry allowed; in this case, the DSL represents a critical point separating ordered phases. On the other hand, if there are no relevant monopole operators that are symmetry allowed, then the DSL is a stable phase of matter --- a gapless spin liquid. However, if there is explicit or spontaneous breaking of a symmetry which was previously forbidding some monopole operator from condensing, this would also lead to a transition to an ordered state.
   
\end{enumerate}

\subsection{Monopole condensation and unbroken symmetries}
\label{sec:monordpar}
The goal of this section is to highlight one fact that will play a crucial role in our work --- in a $120^\circ$ AFM, applying spin rotation about a certain axis on the condensed monopoles is equivalent to applying a $U(1)_{\text{top}}$ phase rotation.\footnote{The reader can skip to Sec.~\ref{sec:josephson}, and return to this section when required.} To do so, we will review a particular mechanism for driving monopole condensation. We first summarize the key facts:
\begin{enumerate}
     \item Under this mechanism, a phase transition occurs when a certain linear combination of $2\pi$ monopole operators that is an eigenvector of a specific $SO(6)$ generator condenses.
     \item From the $G_{IR}$ transformation properties of the condensed monopole operator, one can determine which ordered phase arises.
     \item The $G_{IR}$ symmetry is not fully broken in the ordered state. 
\end{enumerate}

In Appendix~\ref{app:stabilityreview}, we review previous works on the stability of DSLs, which suggest that a DSL could be a stable phase on the triangular lattice. In this case, $2\pi$ monopole operators are symmetry-disallowed in the Langrangian. However, they are still the operators with the lowest scaling dimensions in the low-energy theory. The following mechanism was proposed \cite{song2019unifying,dupuis2019transition,lu2017unification} for destroying the DSL by proliferating $2\pi$ monopoles. First, due to interactions, a fermion bilinear, or mass term, spontaneously acquires a nonzero expectation value. Although other possibilities can occur, we will generally assume that a single bilinear
\begin{equation}
    \langle\bar{\psi}\sigma^{\alpha}\tau^{\beta}\psi\rangle \neq 0
\end{equation}
where $\sigma^\alpha$ and $\tau^\beta$ are Pauli matrices acting in the spin and valley spaces respectively (here $\alpha,\beta \in \{0,1,2,3\}$ but $\alpha$ and $\beta$ are not both 0). This fermion bilinear serves as a mass term that splits the degeneracy of the fermionic Landau zero modes associated with adding a $2\pi$ magnetic flux, lowering the energy (and hence scaling dimension) of one specific linear combination of monopole operators. In particular, given a choice of generator $\sigma^{\alpha}\tau^{\beta}$ of $SU(4)$, we can find a corresponding generator $T\bqty{\sigma^\alpha \tau^\beta}$ of $SO(6)$. This linear combination of monopoles then becomes the most relevant operator and condenses, i.e., acquires a nonzero expectation value. The resulting state is an ordered state in which the fermions are confined~\cite{polyakov1977quark} and the condensed monopole operator serves as an order parameter.

The linear combination of monopoles which condenses corresponds to the eigenvector of $T\bqty{\sigma^\alpha \tau^\beta}$ with the largest eigenvalue. As an example, suppose that the bilinear $\bar{\psi}\sigma^3\psi$ has a nonzero expectation value. One can check that three monopole operators, which we shall call $\hat{\Phi}_{1,2,3}$, transform as $SO(3)_{\text{spin}}$ singlets, and the other three transform as an $SO(3)_{\text{spin}}$ triplet $\hat{\va{\Phi}}$. In this basis, the $SO(6)$ generator corresponding to $\sigma^3$ is
\begin{equation}
    T\bqty{\sigma^3}=   \begin{pmatrix}
                            0_{3\times 3} & \dots & & \\
                            \vdots & 0 & -i &0\\
                            & i & 0 & 0\\
                            & 0 & 0 & 0
                        \end{pmatrix}.
\end{equation}
 The eigenvector of $T\bqty{\sigma^3}$ with maximal eigenvalue (equal to 1) is
\begin{equation}
\begin{aligned}
\label{eq:monexample}
    \expval{\vb{\hat{\Phi}}}&=\begin{pmatrix}\expval{\hat{\Phi}_1}& \expval{\hat{\Phi}_2}& \expval{\hat{\Phi}_3}& \expval{\hat{\Phi}_4}& \expval{\hat{\Phi}_5}& \expval{\hat{\Phi}_6}\end{pmatrix}^T\\&=\abs{\Phi}\begin{pmatrix}0 & 0 & 0 & 1 & i & 0\end{pmatrix}^T.
    \end{aligned}
\end{equation}
The interpretation of this fact is that the fermion mass term makes the energy of the monopole $(\hat{\Phi}_4^\dagger +i\hat{\Phi}_5^\dagger)\ket{GS}$ negative, where $\ket{GS}$ is the ground state of the DSL in the absence of the fermion mass term (whose energy we set to 0). Accordingly, the DSL ground state becomes unstable and a transition occurs to a state with $\expval{\hat{\Phi}_4^\dagger +i\hat{\Phi}_5^\dagger}\neq 0$. 

Although the fermion mass terms pick up nonzero expectation values, the fact that the monopole operators have a lower scaling dimension means that we should treat the condensed monopole operator as the order parameter. Different approaches can be used to determine a microscopic order parameter corresponding to each monopole operator. Refs.~\cite{song2019unifying,song2020spinon} used a symmetry analysis combined with a Wanner center study of mean-field free fermion bands. In Appendix~\ref{sec:monopole}, we combine symmetry analysis with operator algebra constraints to independently motivate the same results. The key results are as follows. First suppose that a spin-triplet monopole operator condenses. If $\expval{\hat{\va{\Phi}}}$ (the vector notation refers to a vector under $SO(3)_{\text{spin}})$ is given by the eigenvector with positive eigenvalue of $\va{d}_{\text{spin}}\cdot\va{T}_{\text{spin}}\bqty{\sigma}$ for a unit 3-vector in spin-space, $\va{d}_{\text{spin}}$, then the ordered state is a $120^\circ$ coplanar AFM order in the plane (in spin-space) normal to $\va{d}_{\text{spin}}$. Similarly, various $\sqrt{12}\times \sqrt{12}$ VBS phases can be obtained if the condensed spin-singlet monopole is an eigenvector of  $\va{d}_{\text{valley}}\cdot \va{T}_{\text{valley}}\bqty{\tau}$ for some unit 3-vector $\va{d}_{\text{valley}}$ in valley-space. In this VBS phase, the area of the unit cell is 12 times the area of the unit cell of a triangular lattice. \footnote{One could also consider condensation channels that are eigenvectors of the mixed generators $T\bqty{\sigma^i \tau^j}$. These ``unconventional orders" were considered in Suppl. Note.~5 of~\cite{song2019unifying}. We will not consider these in this paper.}

Having identified the monopole order parameters, we now notice that the $G_{IR}$ symmetry is not completely broken. Suppose that the condensed monopole is an eigenvector (in the sense of Eq.~\ref{eq:monexample}) of a generator $\hat{Q}$ of $SO(6)$. We focus for later use on the case where $\hat{Q} \in SO(3)_{\text{spin}}$. Then
    \begin{equation}
\label{eq:Szphi}
    \expval{e^{i\theta \hat{Q}}\ \vb{\hat{\Phi}}\ e^{-i\theta \hat{Q}}}=e^{-i\theta Q}\expval{\vb{\hat{\Phi}}}=e^{-i\theta} \expval{\vb{\hat{\Phi}}}.
\end{equation}
Note also that under a $U(1)_{\text{top}}$ phase rotation,
\begin{equation}
\label{eq:bphi}
    \expval{e^{-i\theta \tot{\hat{b}}}\ \vb{\hat{\Phi}}\ e^{i\theta \tot{\hat{b}}}}=e^{i\theta} \expval{\vb{\hat{\Phi}}}.
\end{equation}
Hence $\expval{\vb{\hat{\Phi}}}$ is invariant under $e^{-i \theta \tot{\hat{b}}} e^{i \theta \hat{Q}}$. Such a transformation generates a $SO(2)$ diagonal subgroup of $SO(2)_{\text{spin}}\times U(1)_{\text{top}}$ that is an unbroken symmetry. This ``redundancy'' between spin rotations and $U(1)_{\text{top}}$ phase rotation in the 120$^\circ$ AFM state will play a crucial role in the AC Josephson setup proposed in Section~\ref{sec:josephson}.

For completeness, we mention the concrete connection between the $120^\circ$ AFM order parameter and $\vec{\Phi}$:
\begin{equation}
\label{eq:tripletafm}
    \hat{\va{\Phi}}=\sum_{\vec{n}}e^{-i\vec{Q}.\vec{n}} \pqty{\Sof{\vec{n}}+\ldots},
\end{equation}
where $\vec{Q}=\frac{2\pi}{3}(\vec{\mathrm{b}}_1-\vec{\mathrm{b}}_2)$. Here $\vec{\mathrm{b}}_1\equiv \frac{\sqrt{3}}{2}\hat{x}-\frac{1}{2}\hat{y}$ and $\vec{\mathrm{b}}_2 \equiv \hat{y}$ are reciprocal lattice vectors satisfying $\vec{\mathrm{a}}_i \cdot \vec{\mathrm{b}}_2=\delta_{ij}$, where $\vec{\mathrm{a}}_1\equiv \hat{x},\vec{\mathrm{a}}_2\equiv \frac{1}{2}\hat{x}+\frac{\sqrt{3}}{2}\hat{y}$ are the basis vectors for the triangular lattice. In Eq.~\ref{eq:tripletafm}, the ``$\ldots$'' refers to operators supported on three or more sites.

From Eq.~\ref{eq:tripletafm}, we can see that the ordering pattern
\begin{equation}
\label{eq:afmconfig}
    \expval{\Sof{\vec{n}}}=\begin{pmatrix}\cos \qty(\vec{Q}.\vec{n}),&-\sin \qty(\vec{Q}.\vec{n}),&0\end{pmatrix},
\end{equation}
corresponds to $\expval{\hat{\va{\Phi}}}=\abs{\Phi}\begin{pmatrix}1 & i & 0\end{pmatrix}^T$, which was the example we considered in Eq.~\ref{eq:monexample}.

\section{Monopole Josephson effects}
\label{sec:josephson}
In the previous sections we reviewed how various ordered states can be obtained from a DSL by condensing combinations of 6 monopole operators related to each other by the enlarged $(SO(6)\times U(1)_{\text{top}})/\mathbb{Z}_2$ symmetry. Now we will argue how this can have physical consequences in the form of ``monopole Josephson effects'', by which we mean a flow of monopole current between two symmetry-broken regions of a system. Consider the setup shown in Fig.~\ref{fig:schematic1} where a lattice is split into three regions --- two ordered phases ($O_L$ and $O_R$) separated by a DSL region in the middle. Instead of considering three different materials kept next to each other, we assume that within the \textit{same} sample, perturbations localized to regions $L$ and $R$ drive those regions to ordered phases. This allows us to view the ordered states $O_L$ and $O_R$ as being obtained via monopole condensation from the DSL. Monopoles can now tunnel from $O_L$ to $O_R$ through the middle DSL resulting in a monopole current.

We note that the net monopole current (, i.e., current of $U(1)_{\text{top}}$ charge) is just the emergent electric field rotated by $90^\circ$. This is because Faraday's law takes the form of a conservation law in 2+1 D:
\begin{equation}
    \partial_t \frac{\hat{b}}{2\pi}+\partial_i \frac{\epsilon_{ij} \hat{e}_j}{2\pi}=0.
\end{equation}
So, $\hat{Q}\bqty{{U(1)_{\text{top}}}}(x)=\frac{\hat{b}(x)}{2\pi}$ and $\hat{J}^i\bqty{U(1)_{\text{top}}}(x)=\frac{\epsilon_{ij} \hat{e}_j (x)}{2\pi}$. Therefore this monopole Josephson setup provides a way to induce an emergent electric field inside the DSL (see Fig.~\ref{fig:schematic1}). We will show in Sec.~\ref{sec:raman} that if this induced emergent electric field is time-dependent, it can be optically detected via Raman scattering.

For a given configuration of $O_L$ and $O_R$, our goal is to make predictions for the resulting monopole currents. Since the monopoles are charged under $G_{IR}=SO(6)\times U(1)_{\text{top}}/ \Z_2$, there are 16 different conserved currents in principle, corresponding to each generator of $G_{IR}$. These are $\hat{\vec{J}}\bqty{U(1)_{\text{top}}}$ (analogous to electric Josephson current across superconductors), and 15 currents for each generator of $SO(6)$, of which 3 are spin currents. Deep inside the DSL, since $G_{IR}$ is a symmetry, all 16 currents are conserved at low energies. However, outside the DSL and at the boundaries, generically only the 3 spin currents will be conserved (assuming that $SO(3)_{\text{spin}}$ is respected throughout the system). So we will make statements about two kinds of quantities --- (1) Spin currents that can be measured in either the ordered phases or the DSL, for example using techniques proposed in~\cite{chatterjee2015probing,chen2013detection}, and (2) Currents corresponding to the emergent symmetries of the DSL, which can be probed only within the DSL. The most interesting result of this work is a time dependent (AC) $U(1)_{\text{top}}$ current in the DSL arising due to either a gradient in Zeeman field or due to a gradient in staggered spin chirality applied across the junction. 

In this work, we focus on qualitatively determining, for a given configuration of $O_L$ and $O_R$ and external fields, which monopole currents are non-zero and their dependence on the external fields. In principle, one might also want to calculate the way that the magnitude of the currents $\abs{\expval{\hat{\vec{J}}_{\text{Josephson}}(\vec{x})}}$ scale with the width of the DSL region and thickness of the boundaries. Qualitatively, we expect the currents to decay as a power law in the width $w$ of the DSL region since the DSL is a critical phase. Since the scaling dimension of a conserved current is $d$ in a $d+1$ dimensional CFT, we expect
\begin{equation}
\label{eq:Jmag}
    \abs{\expval{\hat{\vec{J}}_{\text{Josephson}}(\vec{x})}} \propto \frac{\abs{\expval{\vb{\hat{\Phi}}_L}}\abs{\expval{\vb{\hat{\Phi}}_R}}}{w^{2-\Delta_{b,L}-\Delta_{b,R}}}\equiv \mathcal{E}, 
\end{equation}
where $\Delta_{b,L}$ and $\Delta_{b,R}$ are the boundary scaling dimensions of the monopole operators on the left and right boundaries respectively. (Here, we have also defined the RHS of Eq.~\eqref{eq:Jmag} as $\mathcal{E}$ for later convenience.) If the details of the interface provide an additional length scale, this could modify the above scaling. Calculating $\Delta_{b,L}$, $\Delta_{b,R}$ and any additional interface effects is a complicated boundary CFT problem beyond the scope of this work, so we will not address this issue in any more quantitative detail. 

\subsection{Effective Hamiltonian}
\begin{figure}[h!]
    \centering
    \includegraphics[width=0.49\textwidth]{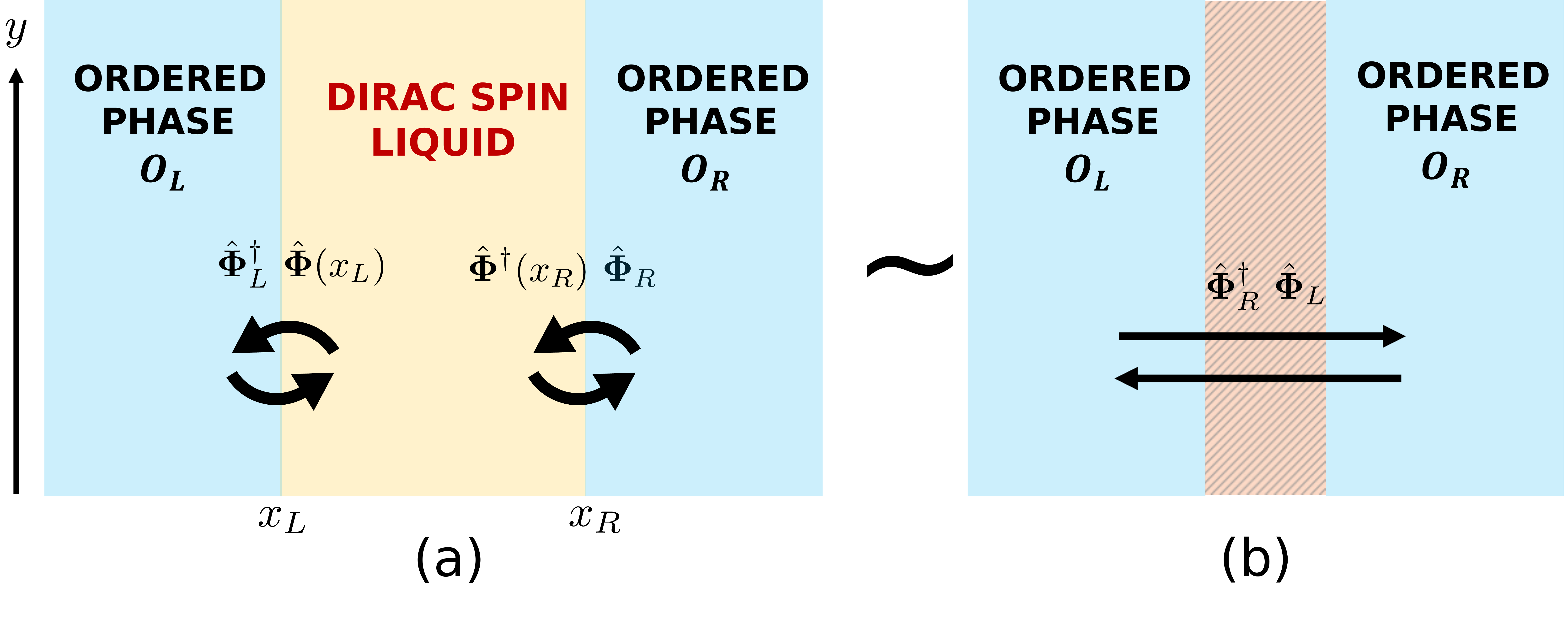}
    \caption{Schematic of the (a) effective monopole tunneling Hamiltonian Eq.~\eqref{eq:hamcoupling} and (b) our proxy Hamiltonian Eq.~\eqref{eq:effham1} used to capture qualitative features of the Josephson currents obtained by schematically ``integrating out'' the DSL.}
    \label{fig:jjLR}
\end{figure}

Our first task is to write a low energy Hamiltonian coupling the two ordered regions $L$ and $R$ to the DSL. In the DSL, monopole operators are the most relevant in the RG sense, and hence coupling terms involving monopole tunneling should be the most important at low energy (we expect this from the large $N_f$ scaling dimensions of monopole operators when one sets $N_f=4$; see Table~\ref{tab:scaldim}). This motivates the following coupling Hamiltonian (see Fig.~\ref{fig:jjLR}(a))
\begin{align}
\label{eq:hamcoupling}
    H_{\text{c}}=&- \sum_{i,j=1}^6 \left(\Gamma_{ij,L} \int d y\  \hat{\Phi}_{iL}^\dagger (x_L,y) \hat{\Phi}_{jD}(x_L,y)\right.  \nonumber \\
    &\left.+ \Gamma_{ij,R}\int dy\  \hat{\Phi}_{iD}^\dagger (x_R,y) \hat{\Phi}_{jR}(x_R,y)\right)+\text{h.c.},
\end{align}
where the left (right) interface is at $x=x_L$ $(x=x_R)$ and $y$ runs parallel to the boundary (in both terms above). A remark on notation --- to emphasize a monopole-tunneling interpretation, we have used the same symbol $\hat{\Phi}_i$ for both the monopole operator in the DSL side ($\hat{\Phi}_{iD}$) and on the ordered sides $\hat{\Phi}_{i,L/R}$. But we note that in general, they would have different scaling dimensions. For example, as one crosses the interface, the system goes through a phase transition and the monopole scaling dimension at the transition is known to be smaller at the phase transition than deep in the DSL (from a large $N_f$ calculation~\cite{dupuis2019transition,zerf2019critical}). 

Since we assumed that the coupling matrix $\Gamma_{ij,L/R}$ preserves spin rotation symmetry,
\begin{equation}
    \Gamma_{ij, L/R}\equiv\Gamma_S \delta_{ij}, \text{ for }4\leq i,j \leq 6.
\end{equation}
Now, since the boundary breaks spatial symmetries, but preserves spin-rotation symmetries, we are also allowed to add single monopole terms for the spin-singlet monopoles, but not spin triplet monopoles:
\begin{equation}
\label{eq:hsource}
    \hat{H}_{\text{source}}=\sum_{i=1}^3 \int dy\pqty{V_{i,L}  \hat{\Phi}_{i}(x_L,y)+V_{i,R} \hat{\Phi}_{i}(x_R,y)}+\text{h.c.} 
\end{equation}
We argue in Appendix~\ref{app:ignore} that $\hat{H}_{\text{source}}$ does not contribute significantly to the currents we are interested in. Then the full Hamiltonian without $\hat{H}_{\text{source}}$, i.e.,   $\hat{H}=\hat{H}_{DSL}+\hat{H}_{O_L}+\hat{H}_{O_R}+\hat{H}_{c}$ has a global spin rotation $SO(3)$ symmetry, and formally, also a global $U(1)_{\text{top}}$ symmetry. We will now use the following strategy --- we first write a schematic Hamiltonian for monopole tunneling between $O_L$ and $O_R$, where the DSL region is assumed to have been ``integrated out'' (we ignore any potential retardation effects coming from integrating out gapless modes in the DSL): 
\begin{equation}
\begin{aligned}
\label{eq:effham1}
    \hat{H}_{\text{eff}}&=\hat{H}_{O_L} + \hat{H}_{O_R} -\Gamma^{\text{eff}}_S \sum_{i=4}^6\pqty{\hat{\Phi}_{iL}^\dagger \hat{\Phi}_{iR} + \text{ h.c.}}\\&-\sum_{i,j=1}^3  \pqty{\Gamma^{\text{eff}}_{ij}\hat{\Phi}_{iL}^{\dagger} \hat{\Phi}_{jR} + \text{ h.c.}},
    \end{aligned}
\end{equation}
where we have neglected the spatial dependence of $\hat{\Phi}_{iL}$ (see Fig.~\ref{fig:jjLR} (b)). The parameter 
\begin{equation}
    \Gamma^{\text{eff}}_S\sim \mathcal{E}L,
\end{equation}
where the factor of $L$, the system length along $y$, comes from the integration along the $y$ direction and $\mathcal{E}$ was estimated in Eq.~\eqref{eq:Jmag}.
We compute the conserved spin current and $U(1)_{\text{top}}$ current flowing from $O_L$ to $O_R$ using the above Hamiltonian, and assume that by current conservation (justified in Appendix~\ref{app:ignore}), the \textit{same} current also flows through the spin liquid. 
\subsection{Brief review of the generalized Josephson effects}
Eq.~\eqref{eq:effham1} is of the form which is usually used to derive generalized Josephson currents between ordered phases which break symmetries belonging to a continuous group $G$~\cite{beekman2020theory,esposito2007field,nitta2015josephson}. Here, we provide a brief review of this formalism which computes the DC and AC Josephson currents of a given symmetry generator, following Ref.~\cite{beekman2020theory}. Let $\hat{Q}^r$ for $r\in \{1,\ldots M\}$ be quantum operators corresponding to the $M$ generators of $G$. For our problem, $\hat{Q}^r$ are the 15 $SO(6)$ charges $\tot{\hat{Q}}\bqty{\sigma^\alpha \tau^\beta}$ (where $\alpha$ and $\beta$ are not both 0) and the emergent flux $\tot{\hat{b}}$. Now suppose the system is divided into left and right parts $L$ and $R$ (Fig.~(\ref{fig:jjLR})). We assume that each $\hat{Q}^r$ can be written as a sum of local operators (see Appendix~\ref{sec:micro} for a discussion). This allows us to define $\hat{Q}^r_{L/R}$ as the restriction of $\hat{Q}^r$ to the respective region $L/R$. Let $\hat{\Phi}_i$ be $N$ operators charged under $G$, i.e. they transform under the group action. The group action is 
\begin{equation}
\label{eq:gaction}
    \comm{\hat{Q}^r}{\hat{\Phi}_i}=\sum_{j=1}^N -T^r_{ij}\hat{\Phi}_j,
\end{equation}
where $T^r$ is an $N\times N$ Hermitian matrix of c-numbers and is a representation of $\hat{Q}^r$ on $\mathbb{C}^N$. When $r$ above corresponds to $U(1)_{\text{top}}$, $T\bqty{U(1)_{\text{top}}}=\mathbb{1}_{6\times 6}$. For explicit formulas for $T^r$ when $G=SO(6)$, see Eq.~\eqref{eq:Tdef}.

The set of operators $\hat{\Phi}_i$ will serve as the order parameter. They will acquire expectation value when the symmetry is broken. Assuming that $\hat{\Phi}_i$ is a sum of local operators, Eq.~(\ref{eq:gaction}) holds approximately even when the operators are restricted to small regions. Suppose the expectation value $\expval{\hat{\Phi}_{iR}}$ on the right differs from that on the left $\expval{\hat{\Phi}_{iL}}$. We now compute the current for each generator $\hat{Q}^r$ from left to right. 

To do this, let us write an effective Hamiltonian. It is identical to Eq.~\eqref{eq:effham1}, except that the following Hamiltonian assumes that the coupling respects the full symmetry $G$:
\begin{equation}
\label{eq:gjjham}
    \hat{H}=\hat{H}_L + \hat{H}_R -\Gamma \sum_{i=1}^N\pqty{\hat{\Phi}_{iL}^\dagger \hat{\Phi}_{iR} + \hat{\Phi}_{iR}^\dagger \hat{\Phi}_{iL}},
\end{equation}
where $\Gamma$ is an effective coupling constant depending on the details of the intermediate region between $L$ and $R$. The current of generator $r$ from left to right $\hat{I}^r_{L\rightarrow R}$ can be calculated from the Heisenberg equation of motion
\begin{equation}
    \hat{I}^r_{L\rightarrow R}\equiv -\dv{\hat{Q}_L^r}{t}=i\comm{\hat{Q}^r_L}{\hat{H}}.
\end{equation}
$Q^r_L$ commutes with $H_L$ because it is conserved, and with $H_R$ because $H_R$ has support only on side $R$. The only nonzero contribution comes from the coupling term.
\begin{equation}
    \hat{I}^r_{L\rightarrow R}=-i\Gamma\sum_{i,j}\hat{\Phi}^\dagger_{iL}T^r_{ij}\hat{\Phi}_{jR}+\text{ h.c.}.
\end{equation}
The expectation value of the RHS above has a disconnected component and a connected component. Since the two sides of the system are symmetry breaking, $\expval{\hat{\Phi}_{jL/R}}$ is macroscopic. So, to lowest order, we will ignore the connected piece. Thus,
\begin{equation}
\label{eq:dcformula}
    \expval{\hat{I}^r_{L\rightarrow R}}\approx-i\Gamma\sum_{i,j}\expval{\hat{\Phi}^\dagger_{iL}}T^r_{ij}\expval{\hat{\Phi}_{jR}}+\text{ h.c.}.
\end{equation}
This is the DC Josephson effect. The same formula can also be used for the AC Josephson effect as follows. A term is added to the Hamiltonian that couples to the \textit{difference} in a conserved charge across the two sides: $\hat{H}_{\mu}=\frac{\mu}{2}(\hat{Q}^s_R -\hat{Q}^s_L)$ (for example, the electric potential difference between the two superconductors). As we will see below, this results in an oscillatory time dependence for $\expval{\hat{\Phi}_{i(L/R)}}$, and therefore according to Eq.~(\ref{eq:dcformula}), the current $\hat{I}^r_{L\rightarrow R}$ also acquires an oscillatory time dependence, 
\begin{align}
    \dv{\hat{\Phi}_{iR}}{t}=-i\frac{\mu}{2}\comm{\hat{\Phi}_{iR}}{\hat{Q}^s_R}&=+i\frac{\mu}{2} \sum_j (T^s)_{ij}\hat{\Phi}_{jR},\\
    \dv{\hat{\Phi}_{iL}}{t}=+i\frac{\mu}{2}\comm{\hat{\Phi}_{iL}}{\hat{Q}^s_L}&=-i\frac{\mu}{2} \sum_j (T^s)_{ij}\hat{\Phi}_{jL}.
\end{align}
The solution is (suppressing the indices of $\vb{\hat{\Phi}}_{L/R}$ and $T^r$)
\begin{equation}
    \hat{\vb{\Phi}}_{R}(t)=e^{i \frac{\mu}{2} t T^s}\hat{\vb{\Phi}}_{R}(0) \text{ and }\hat{\vb{\Phi}}_{L}(t)=e^{-i \frac{\mu}{2} t T^s}\hat{\vb{\Phi}}(0).
\end{equation}
Substituting in Eq.~(\ref{eq:dcformula}), we get 
\begin{equation}
    \label{eq:acformula}
    \begin{aligned}
    \expval{\hat{I}^r_{L\rightarrow R}(t)}\approx &-i\Gamma \expval{\hat{\vb{\Phi}}^\dagger_{L}(0)}e^{i\frac{\mu}{2} t T^s} T^r e^{i\frac{\mu}{2} t T^s}\expval{\hat{\vb{\Phi}}_{R}(0)}\\ &+\text{ h.c.}
    \end{aligned}
\end{equation}
If $T^r$ commutes with $T^s$, then from Eq.~(\ref{eq:acformula}), the current oscillates at frequency $\mu$ --- the familiar AC Josephson effect.

\subsection{Monopole Josephson currents in a DSL}
We will now use the above framework to qualitatively determine the Josephson currents in the setup shown in Fig.~\ref{fig:schematic1}. The symmetry generators $\hat{Q}^r$ are the total magnetic flux $\tot{\hat{b}}$ and the 15 generators of $SO(6)$, namely $\tot{\hat{Q}}\bqty{\sigma^\alpha \tau^\beta}$ (where $\alpha$ and $\beta$ are not both 0. Note that for $\beta=0$ and $\alpha \in \{1,2,3\}$, $\tot{\hat{Q}}\bqty{\sigma^\alpha}$ is just the total conserved spin $\tot{\hat{S}}^{\alpha}$). The operators charged under $\hat{Q}_r$ are the 6 monopoles $\hat{\Phi}_i$ which transform as an $SO(6)$ vector. In Sec.~\ref{sec:monordpar}, we saw that the monopoles serve as order parameters for $120^\circ$ AFMs and $\sqrt{12}\times \sqrt{12}$ valence bond solids. Now consider the scenario shown in Fig.~\ref{fig:schematic1}. Deep inside the DSL, we assume that $G_{\text{space}}$ of the triangular lattice is obeyed. Therefore, $\expval{\hat{\vb{\Phi}}}=0$ here. At the same time, deep inside $O_L$ and $O_R$, monopoles are condensed and acquire macroscopic expectation value $\expval{\hat{\vb{\Phi}}_L}$ and $\expval{\hat{\vb{\Phi}}_R}$ respectively. These ordered phases act as a source of monopoles which can tunnel through the DSL. We show below how one can get DC and AC Josephson effects for the setup where both $O_L$ and $O_R$ are in the $120^\circ$ AFM phase.
\subsubsection{$O_L=120^\circ$ AFM, $O_R=120^\circ$ AFM with angle mismatch: DC Josephson effect}
\begin{figure}[h!]
    \centering
    \includegraphics[width=0.48\textwidth]{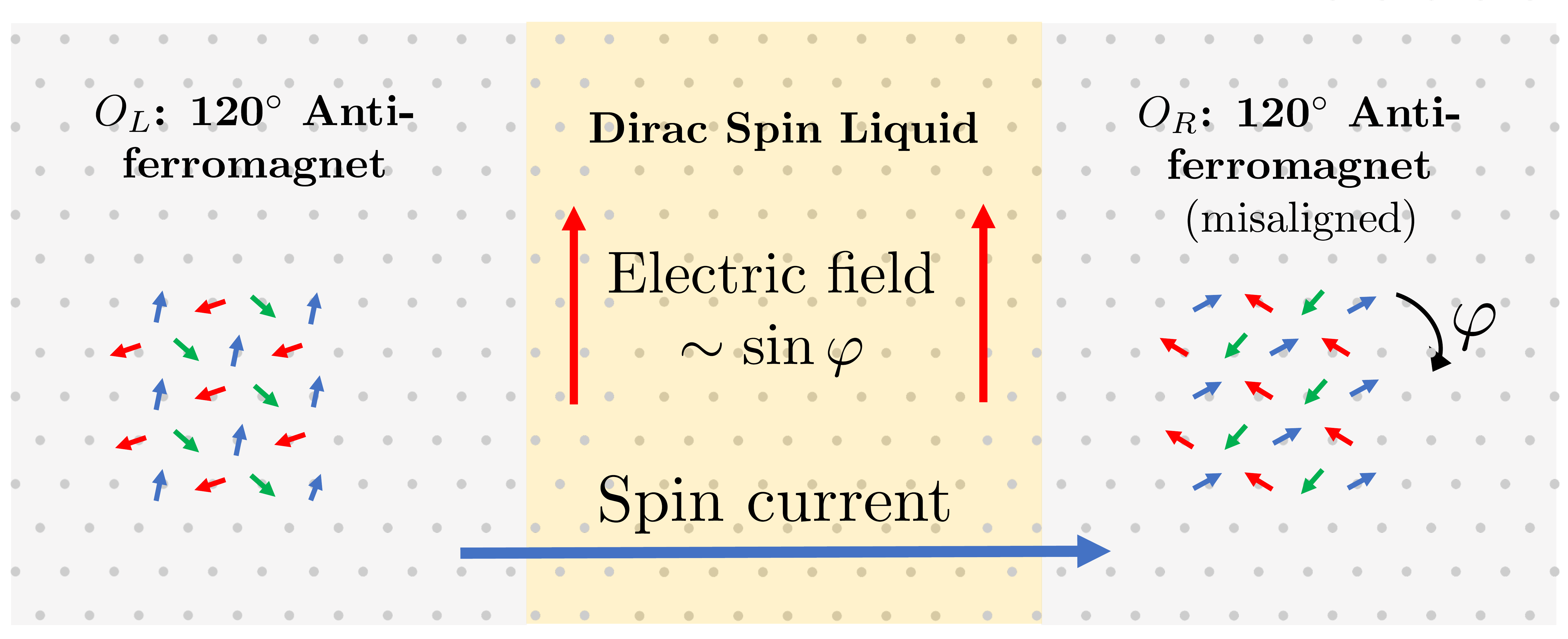}
    \caption{DC Josephson effect: a ($120^\circ$ AFM --- DSL --- $120^\circ$ AFM) arrangement induces a DC electric field inside the DSL. The spins of the $120^\circ$ AFM on the left obey Eq.~(\ref{eq:afmconfig}), while those on the right are rotated with respect to Eq.~(\ref{eq:afmconfig}) by angle $\varphi$. This results in a spin current, whose carriers inside the DSL are monopoles. The resulting monopole current is equivalent to an emergent electric field.}
    \label{fig:dcjos}
\end{figure}
Assume both ordered phases are in a $120^\circ$ AFM state, in which the spin triplet monopoles have acquired nonzero expectation value. Suppose the plane of ordering in spin-space is the same for $O_L$ and $O_R$, which we take to be the $xy$ plane. Now consider the situation where the ordering pattern on $O_R$ is misaligned with respect to $O_L$ by an angle $\varphi$ (see Fig.~\ref{fig:dcjos}). By this, we mean that if the spins in $O_L$ form the ordering pattern given in Eq.~(\ref{eq:afmconfig}), all the spins in $O_R$ are rotated by angle $\varphi$ about the $z-$axis with respect to the configuration dictated by Eq.~(\ref{eq:afmconfig}). (Here, we have assumed that the lattice does not contain any defects.) For this situation, the expectation values of the spin triplet monopoles on either side take the form
\begin{equation}
\label{eq:monconfig1}
    \expval{\hat{\va{\Phi}}_L}=\abs{\Phi_L}\begin{pmatrix}1 &i& 0\end{pmatrix}^T \text{ , } \expval{\hat{\va{\Phi}}_R}=e^{i\varphi}\abs{\Phi_R}\begin{pmatrix}1 &i& 0\end{pmatrix}^T.
\end{equation}
Now, we can apply the formula for Josephson current Eq.~(\ref{eq:dcformula}). Due to the redundancy between $\hat{S}^z$ spin rotation and $U(1)_{\text{top}}$ phase rotation that we observed in Eq.~\eqref{eq:Szphi} and Eq.~\eqref{eq:bphi}, we get both a $U(1)_{\text{top}}$ current and a spin current that are equal to each other:
\begin{equation}
\label{eq:DC120}
    \expval{\hat{I}\bqty{U(1)_{\text{top}}}}=\expval{\hat{I}\bqty{\sigma^3}}=2\Gamma^{\text{eff}}_S \abs{\Phi_L}\abs{\Phi_R} \sin(\varphi).
\end{equation}
Physically, the reason the two currents are the same is that the carriers of conserved spin and the carriers for conserved $U(1)_{\text{top}}$ charge are the same --- the spin triplet monopoles.
The total current $\expval{\hat{I^r}}$ is related to the current density $\abs{\expval{\hat{\vec{J}}^r}}$ as $\abs{\expval{\hat{\vec{J}}^r}}=\expval{\hat{I^r}}/L$ where $L$ is the length of the boundary. These currents are perpendicular to both the $O_L$-DSL and $O_R$-DSL boundaries. Therefore the emergent electric field is parallel to the boundaries. Since $\Gamma^{\text{eff}}_S\sim \mathcal{E}L$, we have 
\begin{equation}
\label{eq:DC120e}
    \expval{\hat{e}}\text{ and }\expval{\hat{J}\bqty{\sigma^3}}\sim\mathcal{E} \sin (\varphi),
\end{equation}
where $\mathcal{E}$ has been estimated in Eq.~\eqref{eq:Jmag}. As we remarked previously, Eq.~\eqref{eq:DC120e} should not be taken quantitatively, hence the $\sim$ symbol. The important takeaway is the $\sin \varphi$ dependence on the angle mismatch and the observation that the Josephson currents are those of the $U(1)_{\text{top}}$ and $\hat{S}^z$ generators, and are in fact equal to each other.
\subsubsection{$O_L=120^\circ$ AFM, $O_R=120^\circ$ AFM: AC Josephson effect}
\label{sec:acjosephson}
We again consider a junction with two $120^\circ$ AFMs separated by a DSL. For this configuration, the expectation values of spin triplet monopole operators on either side of the junction are
\begin{equation}
\label{eq:monconfigac}
    \expval{\hat{\va{\Phi}}_L}=\abs{\Phi_L}\begin{pmatrix}1 &i& 0\end{pmatrix}^T \text{ and } \expval{\hat{\va{\Phi}}_R}=\abs{\Phi_R}\begin{pmatrix}1 &i& 0\end{pmatrix}^T.
\end{equation}
We now propose two scenarios that lead to an AC Josephson effect. The first scenario is analogous to the AC Josephson effect in superconductors obtained by applying a potential difference, a term that couples to the difference in number of particles on the right and left. Analogously, here we can apply the following term to the Hamiltonian that couples to the difference in emergent magnetic flux across the two sides (the conserved $U(1)_{\text{top}}$ charge). On a triangular lattice, we show in Appendix~\ref{sec:charges} that such a term takes the form of the sum of staggered spin chiralities
\begin{equation}
\label{eq:Hmu}
\begin{aligned}
    \hat{H}_{\mu}=&\sum_{\vec{n}}\mu(x)\left(\chiof{\vec{n}}{\vec{n}+\vec{\mathrm{a}}_1}{\vec{n}+\vec{\mathrm{a}}_1-\vec{\mathrm{a}_2}}\right.\\  & \left. -\chiof{\vec{n}}{\vec{n}+\vec{\mathrm{a}}_1}{\vec{n}+\vec{\mathrm{a}}_2}\right)\\ 
    \equiv &\sum_{\vec{n}}\mu(x) \pqty{\downnchi - \upnchi},
    \end{aligned}
\end{equation}
where $\mu (x)$ has a gradient from $L$ to $R$ such that $\mu_R -\mu_L \equiv \mu$. Now, we will use the formula in Eq.~\eqref{eq:acformula} to determine the Josephson currents. In this formula, a perturbation in generator $\hat{Q}^s$ is applied and the current in generator $\hat{Q}^r$ is calculated. For the perturbation considered in Eq.~\eqref{eq:Hmu}, $s$ corresponds to $U(1)_{\text{top}}$. Using Eq.~\eqref{eq:monconfigac} in Eq.~\eqref{eq:acformula}, and noting that $T^s = T\bqty{\tot{\hat{b}}}=\mathbb{1}$ we see that the channels $r$ in which we get nonzero currents are $U(1)_{\text{top}}$ and $\tot{\hat{S}}^z$ (i.e. emergent electric field and spin current). Like before, the two are equal. 
\begin{equation}
    \expval{\hat{I}\bqty{U(1)_{\text{top}}}(t)}= \expval{\hat{I}\bqty{\sigma^3}(t)}=2\Gamma^{\text{eff}}_S \abs{\Phi_L}\abs{\Phi_R} \sin (\mu t).
\end{equation}
Here, we have made use of the observation in Eq.~\eqref{eq:Szphi} and Eq.~\eqref{eq:bphi} that $\expval{\hat{\va{\Phi}}_L}$ and $\expval{\hat{\va{\Phi}}_R}$ are both eigenvectors of $T\bqty{\sigma^3}$ with eigenvalue 1. The above equation says that a difference in spin chirality terms applied across the two ordered phases leads to a time-dependent spin current and an equal emergent electric field. This is a nontrivial prediction of the theory. 

However, applying an external term Eq.~\eqref{eq:Hmu} is not simple experimentally (although there has been a proposal to get a spin chirality term in the effective Floquet Hamiltonian of a spin system driven with a laser~\cite{claassen2017dynamical}). Therefore, we now propose a simpler way to get the same time dependent electric field and spin current as before, but this time exploiting our observation in Eq.~\eqref{eq:Szphi} and Eq.~\eqref{eq:bphi}.

In this second scenario, we apply a Zeeman field gradient across the junction instead of $\hat{H}_\mu$ above, as shown in Fig.~\ref{fig:ramanac},
\begin{equation}
    \hat{H}_h = \sum_{\vec{n}}h(x) \pqty{\hat{S}^z_{\vec{n}}},
\end{equation}
where $h (x)$ has a gradient from $L$ to $R$ such that $h_R -h_L \equiv h$. The only difference now as far as the formula Eq.~\eqref{eq:acformula} is concerned, is that $T^s=T\bqty{\sigma^3}$ instead of $\mathbb{1}$. But since $T\bqty{\sigma^3}\expval{\hat{\va{\Phi}}_{L(R)}}=\expval{\hat{\va{\Phi}}_{L(R)}}$, this difference does not change the currents. We therefore again obtain an AC spin current \textit{and} an equal AC emergent electric field inside the DSL given by
\begin{equation}
\label{eq:unifel}
    \expval{\hat{I}\bqty{U(1)_{\text{top}}}(t)}=\expval{\hat{I}\bqty{\sigma^3}(t)}=2\Gamma^{\text{eff}}_S \abs{\Phi_L}\abs{\Phi_R} \sin (h t).
\end{equation}
We can understand this physically as follows. The presence of the Zeeman field gradient leads to a precession of the macroscopic $120^\circ$ order parameter with a different rate on the two sides of the junction, resulting in a spin current. Similar phenomena have been studied theoretically in several works previously, for conventional magnetically ordered systems~\cite{chasse2010generalized,moor2012josephson,liu2016spin,chen2014dissipationless,ruckriegel2017spin,chandra1990quantum}, and ${}^3$He and spinor BECs~\cite{thuneberg2006theory,qi2009non,misirpashaev1992macroscopic}. 

What is different for our setup is that the proximity to the DSL, and the assumption that the $120^\circ$ AFMs are close to a phase transition to a DSL imply that the carriers of the spin current in the DSLs are monopole operators. Therefore any time-dependent spin current should be accompanied by a time-dependent monopole current, or an emergent electric field in the DSL. 

Describing and probing this electric field is what we will now focus on.
What does an emergent electric field mean in the language of microscopic spins? Using the transformation of electric field under microscopic symmetries (see first row of Table~\ref{tab:currents} in Appendix~\ref{sec:micro}), we can write the following expression for the zero momentum electric field (i.e., integrated over space), keeping only nearest neighbor terms:
\begin{align}
\label{eq:etotal}
    \tot{(\hat{e}_x)}\equiv\int d^2 x \hat{e}_x =&v\sum_{\vec{n}}\pqty{\dotof{\vec{n}}{\vec{n}+\vec{\mathrm{a}}_2}-\dotof{\vec{n}}{\vec{n}+\vec{\mathrm{a}}_2-\vec{\mathrm{a}}_1}}\nonumber \\&+\ldots,\\
    \tot{(\hat{e}_y)}\equiv\int d^2 x \hat{e}_y =&v\sum_{\vec{n}}\frac{1}{\sqrt{3}}\left(2\dotof{\vec{n}}{\vec{n}+\vec{\mathrm{a}}_1}-\dotof{\vec{n}}{\vec{n}+\vec{\mathrm{a}}_2}\right. \nonumber \\ &\left.-\dotof{\vec{n}}{\vec{n} + \vec{\mathrm{a}}_2 - \vec{\mathrm{a}}_1}\right)+\ldots,
\end{align}
where $v$ is a constant of the order of the Dirac velocity, which in turn is of the order of $J\mathrm{a}$ where $J$ is the exchange coupling strength and $\mathrm{a}$ is the lattice spacing. As a consequence of the AC Josephson effect discussed above, we expect 
\begin{equation}
\label{eq:eexpval}
    \pqty{\expval{\hat{e}_x(t)}, \expval{\hat{e}_y(t)}}= \mathcal{E} \sin (ht) \ \pqty{\cos \theta, \sin \theta},
\end{equation}
where $\theta$ is the angle made by the electric field with $x$-axis (the direction of the electric field is tangential to the DSL-AFM boundaries). $\mathcal{E}$ is given by the right-hand side of Eq.~\eqref{eq:Jmag}, calculating which is beyond the scope of this work. The key point is that since the DSL is described by a CFT, $\mathcal{E}$ decays only as a power law in the width of the DSL. 

We see that the operators $\hat{e}_x$ and $\hat{e}_y$ have a non-trivial spatial structure. In order to detect this ``electric field'' consistently, we would need a probe that is sensitive to rotational form-factors. Optical probes are well-suited for this purpose because of the control one gets from the direction of polarization of light~\cite{devereaux2007inelastic,nasu2016fermionic}. We now propose a way to measure the emergent AC electric field inside the DSL using Raman scattering.

\section{Raman scattering probe of emergent electric field}
\label{sec:raman}
\begin{figure}[h!]
    \centering
    \includegraphics[width=0.48\textwidth]{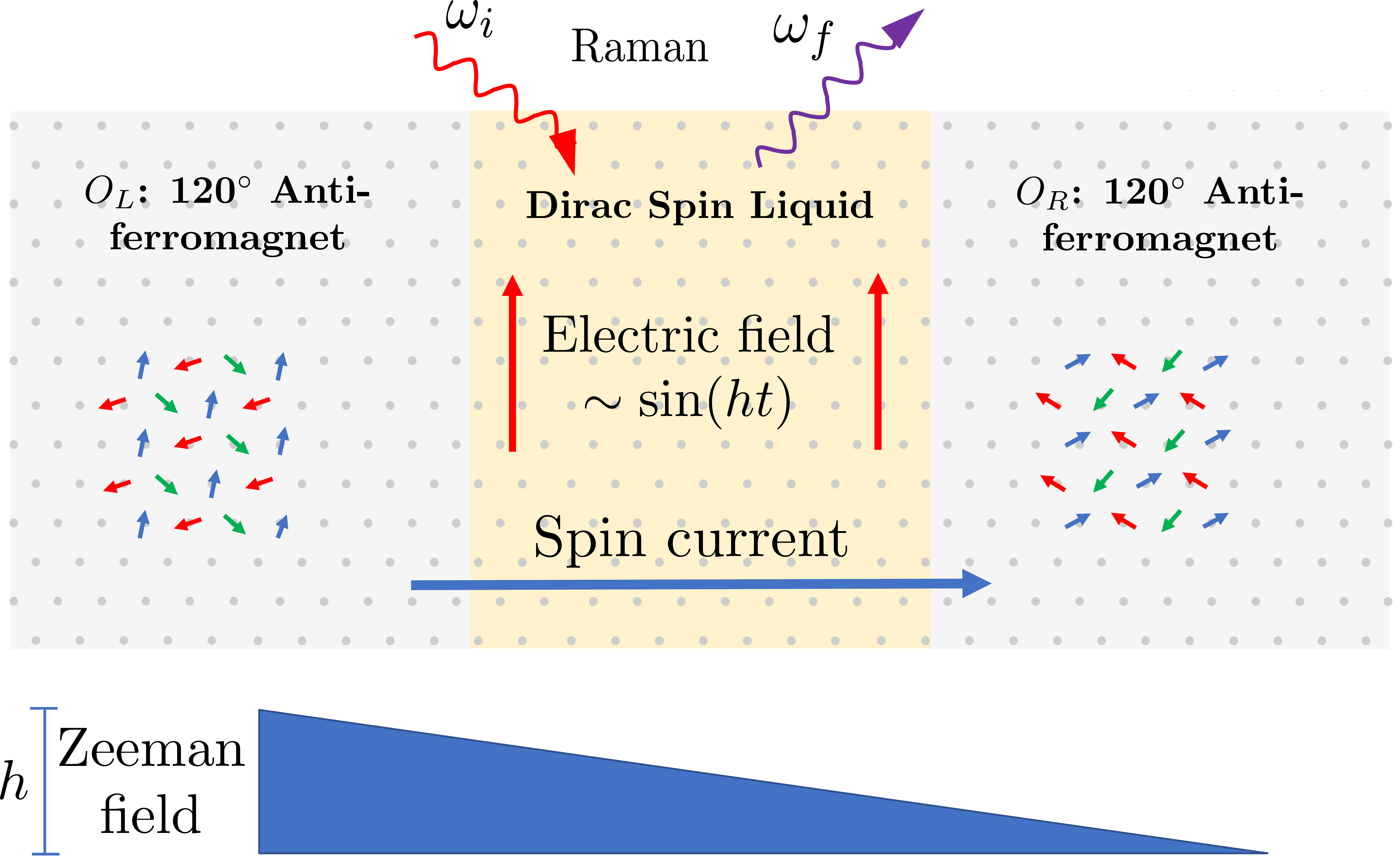}
    \caption{The proposed setup ($120^\circ$ AFM --- DSL --- $120^\circ$ AFM) to induce and probe the AC Josephson effect. An out of plane (w.r.t. magnetic ordering) Zeeman field gradient of magnitude $h$ is applied across the junction, causing the spins on the left to precess at a different rate than the spins on the right. This precession results in a spin current, whose carriers inside the DSL are monopoles. The resulting emergent electric field within the DSL can be probed via Raman scattering.}
    \label{fig:ramanac}
\end{figure}
Suppose the DSL region is irradiated with a laser of frequency $\omega_i$. A Raman signal corresponds to inelastic scattering of light, i.e. the outgoing photon's frequency $\omega_f$ is different from $\omega_i$. We will now argue that the presence of an emergent electric field in the DSL of frequency $h$, and in particular the one produced in the setup considered in Sec.~\ref{sec:acjosephson} (see Fig.~\ref{fig:ramanac}), will lead to peaks at Raman frequency shifts $ \omega_{\Delta}\equiv \omega_f - \omega_i = \pm h$. 

It was shown in~\cite{shastry1990theory,shastry1991raman} that the Raman scattering rate $R$ for a spin system (not necessarily a spin liquid) is given by the following correlation function calculated in an energy eigenstate of the spin system $\ket{i}$
\begin{equation}
\label{eq:scatrate}
    R=\int_{-\infty}^{\infty}dt e^{i \omega_{\Delta}t} \mel{i}{\hat{M}^{\dagger}_{\vec{q}}(0)\hat{M}_{\vec{q}}(t)}{i},
\end{equation}
where $\vec{q}=\vec{q}_f - \vec{q}_i$ is the momentum transferred to the photon. (For simplicity, we will ignore this small momentum transfer from now on.) The operator $\hat{M}$ acts on the Hilbert space of the spin system, and depends on the underlying lattice for the spin system as well as the polarizations and momenta of the incident and scattered light. The operator $\hat{M}$ was calculated for some relevant cases in~\cite{shastry1990theory,ko2010raman}, and we will present the leading-order results later in Eq.~\eqref{eq:fleuryloudon}.

However, if the expectation value of some operator (in our case, the emergent electric field) $\mel{\psi}{\hat{\vec{e}}(t)}{\psi}$ in a state $\ket{\psi}$ were to depend sinusoidally on time, then $\ket{\psi}$ is clearly not an energy eigenstate but rather a nonequilibrium state. In such a state, we show in Appendix~\ref{app:raman} that Eq.~\eqref{eq:scatrate} gets modified and the Raman scattering rate now measures the following time-averaged correlation function of the same operator $\hat{M}$:
\begin{equation}
\label{eq:Rgeneral}
\begin{aligned}
    R=\lim_{T\to \infty}\frac{1}{T}\int_{-\frac{T}{2}}^{\frac{T}{2}}dt_0 \int_{-\frac{T}{2}}^{\frac{T}{2}} dt  \mel{\psi}{\hat{M}^{\dagger}(t_0)\hat{M}(t+t_0)}{\psi}\\
    \times e^{i \omega_{\Delta}t}.
    \end{aligned}
\end{equation}
We will use the following (Fleury-Loudon~\cite{fleury1968scattering}) form for $\hat{M}$,
\begin{equation}
    \label{eq:fleuryloudon}
    \begin{aligned}
        \hat{M}=\sum_{\vec{n},\vec{n}^\prime}\frac{2 \tnew_{\vec{n},\vec{n}'}^2 \tilde{e}^2}{U-\omega_i} \qty{\vec{\epsilon}_f^*\cdot(\vec{n}' -\vec{n})}\qty{\vec{\epsilon}_i \cdot(\vec{n}' - \vec{n})}\\ \times \pqty{\frac{1}{4}-\dotof{\vec{n}}{\vec{n}'}},
        \end{aligned}
\end{equation}
which requires a bit of explanation. Here $\vec{\epsilon}_i$ and $\vec{\epsilon}_f$ are the polarizations of the incoming and outgoing photons, respectively. Eq.~\ref{eq:fleuryloudon} assumes that the spins arise from a single band Hubbard model at half filling in the large $U$ limit, of the following form
\begin{equation}
\begin{aligned}
    \hat{H}_{\text{el}}=&-\pqty{\sum_{\vec{r},\vec{r}^\prime,\sigma}\tnew _{\vec{r},\vec{r}^\prime}\hat{c}_{\vec{r}\sigma}^\dagger \hat{c}_{\vec{r}^\prime\sigma}e^{ie \hat{\vec{A}}\pqty{\frac{\vec{r}+\vec{r}^\prime}{2}}\cdot(\vec{r}-\vec{r}^\prime)}+\text{h.c.}}\\
    &+ U\sum_{\vec{r}}\hat{\mathrm{n}}_{\vec{r},\uparrow}\hat{\mathrm{n}}_{\vec{r},\downarrow}.
    \end{aligned}
\end{equation}
Here $\hat{\vec{A}}(\vec{r})$ is the electromagnetic field and has the following expansion in photon creation and annihilation operators
    \begin{equation}
        \hat{\vec{A}}(\vec{r})=\sum_{\vec{k}}\frac{1}{\sqrt{2\varepsilon V \omega_{\vec{k}}}} \pqty{\vec{\epsilon}_{\vec{k}}\hat{a}_{\vec{k}}+\vec{\epsilon}^{*}_{-\vec{k}}\hat{a}^{\dagger}_{-\vec{k}}}e^{i \vec{k}.\vec{r}},
    \end{equation}
where $\vec{\epsilon}_{\vec{k}}$ is the polarization of mode $\vec{k}$, and $\varepsilon$ and $V$ are the dielectric constant and laser mode volume respectively. We have defined the coupling constant $\tilde{e}^2\equiv \frac{e^2 \mathrm{a}^2 \sqrt{\mathcal{N}_i}}{2\varepsilon V \sqrt{\omega_i \omega_f}} $ where $e$ is the electron charge and $\mathcal{N}_i$ is the initial number of photons in the mode of frequency $\omega_i$. The driving is assumed to be near resonance, but at the same time satisfying $\tnew_{\vb{r}\vb{r}^\prime} \ll \abs{U-\omega_i}\ll U$. Under this assumption, one can calculate the scattering rate perturbatively in both $\tnew/(\omega_i -U)$ and $\tilde{e}$, the light-matter coupling constant; one obtains Eq.~\ref{eq:fleuryloudon} at order $\tilde{e}^2 \tnew ^2 /(\omega_i - U)$. 

It is convenient to decompose the tensor $(\epsilon_f^j)^{*} \epsilon_i^k$ into two 1-dimensional ($A_{1g}, A_{2g})$  and one 2-dimensional ($E_g$) irreducible representations of the triangular lattice point group
\begin{equation}
\begin{aligned}
    A_1 &\equiv (\epsilon_f^x)^*\epsilon_i^x + (\epsilon_f^y)^* \epsilon_i^y,\\
    A_2 &\equiv (\epsilon_f^x)^*\epsilon_i^y - (\epsilon_f^y)^* \epsilon_i^x,\\
    \begin{pmatrix}
    E_1\\
    E_2 
    \end{pmatrix}
    &\equiv 
    \begin{pmatrix}
    -(\epsilon_f^x)^*\epsilon_i^x + (\epsilon_f^y)^* \epsilon_i^y\\
    (\epsilon_f^x)^*\epsilon_i^y + (\epsilon_f^y)^* \epsilon_i^x
    \end{pmatrix}.
\end{aligned}
\end{equation}
On the triangular lattice, using this basis reduces Eq.~\eqref{eq:fleuryloudon} to
    \begin{equation}
    \begin{split}
        \hat{M}=&\frac{4 \tnew^2 \tilde{e}^2}{U-\omega_i}\pqty{A_1 \hat{\mathcal{O}}_{A_1}+E_2 \hat{\mathcal{O}}_{E_2}-E_1 \hat{\mathcal{O}}_{E_1}}, \quad \text{where}\\
        \hat{\mathcal{O}}_{A_1}=&\sum_{\vec{n}}\pqty{\dotof{\vec{n}}{\vec{n}+\vec{\mathrm{a}}_1} + \dotof{\vec{n}}{\vec{n}+\vec{\mathrm{a}}_2} + \dotof{\vec{n}}{\vec{n}+\vec{\mathrm{a}}_2 - \vec{\mathrm{a}}_1}},\\
        \hat{\mathcal{O}}_{E_2}=&\frac{\sqrt{3}}{4}\sum_{\vec{n}}\pqty{\dotof{\vec{n}}{\vec{n}+\vec{\mathrm{a}}_2} - \dotof{\vec{n}}{\vec{n}+\vec{\mathrm{a}}_2 - \vec{\mathrm{a}}_1}},\\
        \hat{\mathcal{O}}_{E_1}=&\frac{1}{4}\sum_{\vec{n}}\left(2\dotof{\vec{n}}{\vec{n}+\vec{\mathrm{a}}_1} - \dotof{\vec{n}}{\vec{n}+\vec{\mathrm{a}}_2} \right. \\
        &\left.- \dotof{\vec{n}}{\vec{n}+\vec{\mathrm{a}}_2 - \vec{\mathrm{a}}_1}\right).
        \end{split}
    \end{equation}
    Up to order $\frac{\tnew^2 \tilde{e}^2}{U-\omega_i}$, there is no term in the $A_{2g}$ channel. For the particular case of a Dirac spin liquid, we can use Eq.~\eqref{eq:etotal} to relate the emergent electric fields to microscopic quantities. Up to corrections involving longer range terms, we see that $\hat{\mathcal{O}}_{E_2}$ and $\hat{\mathcal{O}}_{E_1}$ are indeed proportional to the emergent electric fields $\tot{(\hat{e}_x)}$ and $\tot{(\hat{e}_y)}$ respectively (because the symmetry transformation of the emergent electric fields on the triangular lattice is identical to that of the $E_{2g}$ channel.) On the other hand, $\hat{\mathcal{O}}_{A_1}$ is proportional to the Hamiltonian of the system. This lets us write the above expression as 
\begin{equation}
    \hat{M}=\frac{4\tnew^2 \tilde{e}^2}{U-\omega_i}\pqty{\frac{1}{J} A_1 \hat{H} + \frac{\sqrt{3}\mathcal{A}}{4v}(E_2 \hat{e}_x - E_1 \hat{e}_y)+\ldots},
\end{equation} 
where $\mathcal{A}$ is the area of the DSL region.
We can now relate the Raman scattering rate in a DSL to correlation functions of the electric field and the Hamiltonian by inserting the above expression into Eq.~\eqref{eq:Rgeneral}. 

Before we proceed, we highlight two main differences from previous theoretical literature, arising due to the presence of the AC Josephson effect in the setup in Sec.~\ref{sec:acjosephson},  on Raman scattering. First, the Raman scattering rate is usually derived when the spin system is in an equilibrium state, where one-point functions $\expval{\hat{\mathcal{O}}(t)}$ for interesting operators $\hat{\mathcal{O}}$ typically equal zero, in which case a correlation function $\expval{\hat{\mathcal{O}}_1(t_1)\hat{\mathcal{O}}_2(t_2)}$ would be given entirely by its connected component. However, in our case, the DSL is in a nonequilibrium steady state where $\expval{\hat{\vec{e}}(t)}\propto \sin (h t)$ (see Eq.~\ref{eq:eexpval}). Hence the correlation function also has a disconnected component. In what follows, we will assume that the contribution to the autocorrelation function coming from the monopole Josephson effect is dominated by the disconnected piece
\begin{equation}
    \expval{\hat{e}_i(t_1)\hat{e}_j(t_2)}\approx \expval{\hat{e}_i(t_1)}\expval{\hat{e}_j(t_2)}.
\end{equation}
Since 
\begin{equation}
\begin{aligned}
    \lim_{T\to \infty}\frac{1}{T}\int_{-T/2}^{T/2}d t_0 \sin (h t_0) \sin(h (t+t_0))&=\frac{1}{2}\cos (h t) \text{ and}\\
    \lim_{T\to \infty}\frac{1}{T}\int_{-T/2}^{T/2} d t_0 \sin (h (t+t_0))&=0,
    \end{aligned}
\end{equation}
we find that the autocorrelation functions are sharply peaked in frequency as follows
\begin{equation}
\begin{aligned}
    \lim_{T\to \infty}\frac{1}{T}\int_{-\frac{T}{2}}^{\frac{T}{2}}d t_0 \expval{\hat{e}_x(t_0)\hat{e}_x(t_0 +t)}&\approx\frac{\mathcal{E}^2 \cos^2 \theta}{2}  \cos (ht),\\
    \lim_{T\to \infty}\frac{1}{T}\int_{-\frac{T}{2}}^{\frac{T}{2}}d t_0 \expval{\hat{e}_x(t_0)\hat{e}_y(t_0 +t)}&\approx\frac{\mathcal{E}^2 \sin^2 \theta}{2} \cos (ht),\\
    \label{eq:exey}
    \lim_{T\to \infty}\frac{1}{T}\int_{-\frac{T}{2}}^{\frac{T}{2}}d t_0 \expval{\hat{e}_x(t_0)\hat{e}_y(t_0 +t)}&\approx\frac{\mathcal{E}^2 \sin (2\theta)}{4} \cos (ht),\\
    \lim_{T\to \infty}\frac{1}{T}\int_{-\frac{T}{2}}^{\frac{T}{2}}d t_0 \expval{\hat{e}_x(t_0)\hat{H}(t_0 +t)}&\approx 0.
    \end{aligned}
\end{equation}
This brings us to the second difference --- an equilibrium correlation function in a symmetry preserving state is diagonal in the $A_1$, $A_2$, $E_2$, $E_1$ basis. But due the monopole Josephson effect, the steady state no longer has rotational symmetry, leading to mixing within the $E_g$ channel, (see Eq.~\eqref{eq:exey}). Now, we are ready to write down the final result for the Raman scattering rate
\begin{equation}
\label{eq:ramanprediction}
    R=\mathcal{K}\abs{E_1 \sin \theta - E_2 \cos \theta}^2\qty{\delta(\omega_{\Delta} -h)+\delta( \omega_{\Delta} +h)},
\end{equation}
where $\mathcal{K}\equiv \frac{3\pi}{2} \mathcal{E}^2\pqty{\frac{\mathcal{A}}{v}\frac{\mathrm{t}^2 \tilde{e}^2 }{U-\omega_i}}^2$ is a constant.
By tuning the polarizations of the incoming and detected photons, one can tune the values of $A_1, A_2, E_1, E_2$. By measuring the scattering rate $R$ for each such choice, one can separately measure the correlation function in  each channel, and thus verify the prediction in Eq.~\eqref{eq:ramanprediction}. 

We have shown that for a junction with two $120^\circ$ AFMs separated by a DSL, if we apply a Zeeman field gradient $h$ across the junction, the AC emergent electric field resulting from the monopole Josephson effect produces sharp peaks at Raman frequency shifts $\pm h$ in the $E_g$ channels. The strength of the peak decays as a power law in the width of the junction. The above setup provides a way to induce and directly probe the emergent electric field in a Dirac spin liquid.
\section{Other monopole Josephson effects}
\label{sec:othereff}
In this section, we present other effects which fall under the general umbrella of monopole Josephson effects.
\subsection{Josephson energy --- Long range phase rigidity}
For the DC Josephson current, we assumed that the two $120^\circ$ orders had an angle misalignment. This misalignment could have arisen due to an external pinning potential, with a strength smaller than the coupling $\Gamma^S_{\text{eff}}$, so that our assumption that the spin $SO(3)$ is conserved continues to hold. Here, on the other hand, we suppose that there is no external pinning and we let the relative angle between $\hat{\va{\Phi}}_L$ and $\hat{\va{\Phi}}_R$ to fluctuate. In other words, if $\expval{\va{\Phi}_L} =\abs{\Phi_L} \begin{pmatrix}1&i&0\end{pmatrix}$ and $\expval{\va{\Phi}_R} = e^{i\varphi}\abs{\Phi_R}\begin{pmatrix}1&i&0\end{pmatrix}$, we let $\varphi$ be a dynamical degree of freedom. 

As a first approximation, one can calculate the energy of such a configuration from the Hamiltonian Eq.~(\ref{eq:hamcoupling}) from just the disconnected piece of the two-monopole correlation function
\begin{equation}
\begin{aligned}
    E\bqty{\varphi}&\approx -\mathcal{E}L\pqty{\expval{\hat{\va{\Phi}}^\dagger_L} \vdot \expval{\hat{\va{\Phi}}_R} + \text{ h.c.}}\\ &\sim -2\mathcal{E}L\abs{\Phi_L}\abs{\Phi_R} \cos(\varphi).
    \end{aligned}
\end{equation}
This Josephson energy implies that there is a restoring force that tries to align the angles of two $120^\circ$ AFM puddles separated by a DSL. This restoring force is proportional to $\mathcal{E}$, which we expect to decay only as a power law in the width of the DSL (see Eq.~\eqref{eq:Jmag}) since the DSL is a critical phase. Therefore puddles of ordered phases separated by regions of DSL will display a tendency for their order parameters to align, a behavior we call long-range phase rigidity.
\subsection{Mixed current: $O_L=120^\circ$ AFM, $O_R=$ VBS}
Consider a junction with a DSL region separating two completely different orders: $O_L=120^\circ$ AFM and $O_R$= $\sqrt{12}\times \sqrt{12}$ VBS.  Using the parent state picture, we view $O_L$ as the condensate of spin triplet monopoles and $O_R$ as the condensate of spin singlet monopoles, as follows:
\begin{align}
\label{eq:egmixed}
    \expval{\begin{pmatrix}\Phi_1 & \Phi_2 & \Phi_3 & \Phi_4 & \Phi_5 & \Phi_6\end{pmatrix}}_L^T&=\abs{\Phi_L}\begin{pmatrix}0&0&0&1&i&0\end{pmatrix}^T, \nonumber \\
    \expval{\begin{pmatrix}\Phi_1 & \Phi_2 & \Phi_3 & \Phi_4 & \Phi_5 & \Phi_6\end{pmatrix}}_R^T&=\abs{\Phi_R}\begin{pmatrix}1&i&0&0&0&0\end{pmatrix}^T.
\end{align}
We now argue that in this configuration, the DSL will have currents of the mixed spin-valley generators of $SO(6)$. As a toy model, we start with the coupling Hamiltonian in Eq.~\eqref{eq:gjjham}. This assumes an $SO(6)$ symmetric term for monopole tunneling. In this case, one can directly use Eq.~\eqref{eq:dcformula} to calculate the Josephson currents. We then see that the currents with nonzero expectation value are those for the the mixed $SO(6)$ generators: $\hat{\vec{J}}\bqty{\sigma^i \tau^j}$. For the example we picked in Eq.~\eqref{eq:egmixed}, $\hat{\vec{J}}^{14}\equiv \hat{\vec{J}}\bqty{\sigma^1 \tau^1}$ and $\hat{\vec{J}}^{25}\equiv \hat{\vec{J}}\bqty{\sigma^2 \tau^2}$ are nonzero and equal, and the remaining independent currents are 0. 

The presence of these mixed currents can be identified by their symmetry breaking patterns in the bulk of the DSL region. In Appendix~\ref{sec:micro}, we summarize the symmetry properties of all such currents (Table~\ref{tab:currents}) on the triangular lattice. We observe that when a general combination of the mixed currents (last three rows of Table~\ref{tab:currents}) has a nonzero expectation value, time-reversal symmetry is broken and discrete translation symmetry is reduced to translations by 2 lattice spacings along both $\vec{\mathrm{a}}_1$ and $\vec{\mathrm{a}}_2$, i.e., a 4-site unit cell forms in the DSL region. 

We point out, however, that the microscopic model does not have an $SO(6)$ symmetry. Therefore, our assumption above of an $SO(6)$-symmetric coupling at the interface is not strictly justified. Nevertheless, we can qualitatively argue that the physical consequence of having mixed currents in the DSL --- unit-cell expansion and time-reversal symmetry breaking (within the DSL bulk) continues to hold. Consider the effective Hamiltonian Eq.~\eqref{eq:effham1} with a coupling that breaks $SO(6)$ to $SO(3)_{\text{spin}}$. Then we have
\begin{equation}
    \begin{aligned}
        \expval{-\dot{\hat{Q}}_L\bqty{\sigma^i}}&=\expval{\dot{\hat{Q}}_R\bqty{\sigma^i}}=0 \text{ and }\\
        \expval{-\dot{\hat{Q}}_L\bqty{\tau^i}}&=\expval{\dot{\hat{Q}}_R\bqty{\tau^i}}=0, \text{ while }\\
        \expval{-\dot{\hat{Q}}_L\bqty{\sigma^i \tau^j}}&\neq\expval{\dot{\hat{Q}}_R\bqty{\sigma^i \tau^j}} \text{ but both are nonzero.} 
    \end{aligned}
\end{equation}
The above equation says that, not surprisingly, the charge of mixed generators lost from side $L$ is not equal to that gained by side $R$. This leakage however, is localized to the boundaries, because deep inside the DSL, the mixed currents are still conserved. Generically then, the expectation value of some combination of mixed currents in the DSL should still be nonzero. This is true in the limit of $SO(6)$-symmetric coupling, and as we go away from this point, there is no reason for the mixed currents to immediately drop to zero.    
\subsection{Response to $U(1)_{\text{top}}$ flux insertion --- lattice dislocation}
\begin{figure}[h!]
    \centering
    \includegraphics[width=0.48\textwidth]{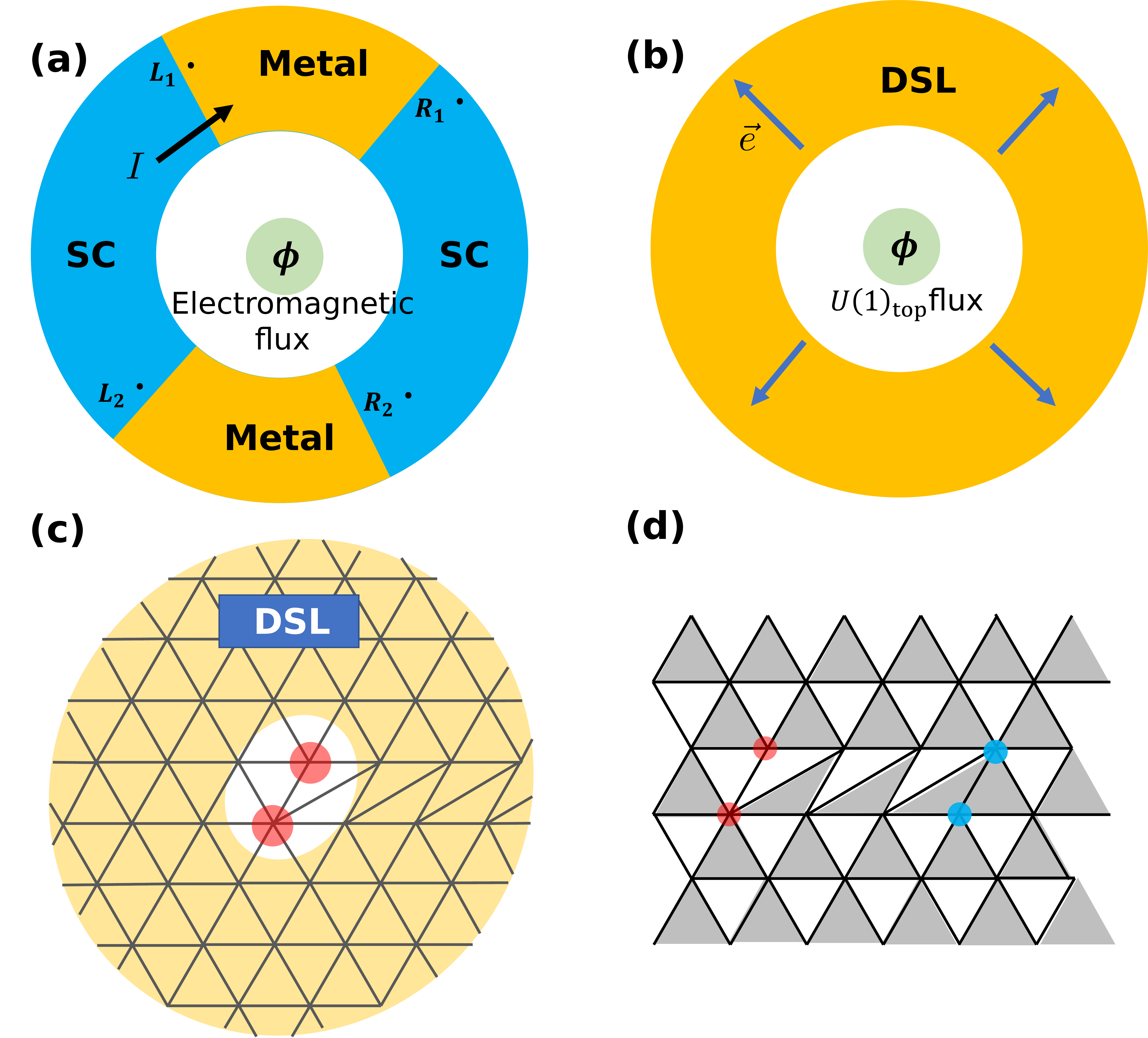}
    \caption{(a): In a SQUID geometry (SC-metal-SC-metal), threading a flux $\phi$ through the center results in a tangential electric current $I$ (black arrow). (b): Similarly, a DSL with a flux $\phi$ in $U(1)_{\text{top}}$ going through results in an emergent electric field $\expval{\hat{\vec{e}}}$ radially outwards (blue arrows). (c): A DSL in the presence of a lattice dislocation. The red circles mark the two lattice sites making up the dislocation. (d): Mean field considered for numerics in the presence of two dislocations with opposite Burger's vectors (red and blue). Grey triangle indicates $\pi$ flux.}
    \label{fig:fluxschem}
\end{figure}

For the familiar DC Josephson effect, a phase difference between the right ($R$) and left ($L$) superconductors is maintained by threading a magnetic flux, like the SQUID geometry shown in Fig.~\ref{fig:fluxschem}(a), because the gauge invariant phase difference between points $R1$ and $L1$ is $\theta_{R1}-\theta_{L1}+e\int_{L1}^{R1}\vec{A}\cdot d\vec{r}$. This results in a current between the two superconductors in the tangential direction.

 In Fig.~\ref{fig:fluxschem}(b), we consider a related configuration involving the DSL. Here $\phi$ is the flux inserted in $U(1)_{\text{top}}$. For simplicity, we have assumed that the full system is in the DSL phase. An analogous situation for metals is persistent currents in the ground state~\cite{bleszynski2009persistent} in the presence of magnetic flux, as required by the Byers-Yang theorem~\cite{byers1961theoretical}:
\begin{equation}
    I\bqty{\phi}=-\frac{1}{T}\frac{\partial F\bqty{\phi}}{\partial \phi},
\end{equation}
where $I$ is the equilibrium current, $T$ is the temperature, and $F$ is the free energy. Similar to an electron current in a SQUID, this results in a tangential $U(1)_{\text{top}}$ current, i.e. a radial electric field. But how do we insert a flux in $U(1)_{\text{top}}$? Lattice translations are known to have a nontrivial $U(1)_{\text{top}}$ action when embedded into the low-energy symmetry group $G_{IR}$~\cite{song2019unifying,song2020spinon}. Therefore, a symmetry defect of lattice translation, that is, a lattice dislocation (see Fig.~\ref{fig:fluxschem}(c)), serves as a $U(1)_{\text{top}}$ flux\footnote{In addition to being a $U(1)_{\text{top}}$ symmetry flux, it is also an $SO(6)$ symmetry flux, but this fact does not play a role in the rest of our discussion.}.

Following the above argument, we expect that a lattice dislocation creates a radial electric field, which by Gauss's Law results in a \textit{spinon} charge ($U(1)$ gauge charge and \textit{not} a $U(1)_{\text{top}}$ charge) near the dislocation. As a first step, in the parton picture, we can verify this prediction at the mean field level. We consider the mean-field ansatz shown in Fig.~\ref{fig:fluxschem}(d) on a lattice with two dislocations with opposite Burger's vectors ($\vec{\mathrm{a}}_2$ and $-\vec{\mathrm{a}}_2$ respectively) separated by $d$ lattice spacings. The triangles shaded grey have $\pi$ flux going through them. This ansatz preserves time-reversal but breaks charge-conjugation symmetry. We then numerically diagonalize the corresponding free fermion Hamiltonian on an $L\times L$ torus. The lowest $L^2/2$ levels are filled by both spin $\uparrow$ and $\downarrow$ fermions in the ground state. Then we compute the charge in a region $D$ enclosing the dislocation
\begin{equation}
    \expval{\hat{q}_{\text{dislo}}}=2\sum_{\vec{r}\in D}\pqty{\sum_{i=1}^{L^2/2}\pqty{\abs{\psi^i_{\vec{r}}}^2-\frac{1}{2}}},
\end{equation}
where $\psi^i_{\vec{r}}$ is the single fermion wavefunction of the $i^{th}$ eigenstate (sorted in increasing order of energy) evaluated at $\vec{r}$. For $L=100$, $d=50$ and $D$ being a circle of radius 20, we get $q_{\text{dislo}}=0.297$.

We are unable to determine whether a non-zero spinon charge survives once we include gauge fluctuations. Qualitatively, we expect that the charge gets renormalized due to screening, but may not drop to 0. If the localized spinon charge is indeed nonzero, what it would mean in terms of microscopic spins is an open question. In Eq.~\eqref{eq:expcharge} of Appendix~\ref{sec:micro}, we have written a nontrivial microscopic operator that is consistent with both the symmetry properties of the field theory spinon charge operator and Gauss's law. How this expression gets modified for a lattice with dislocations, and whether any resulting spinon charge can be computed numerically in candidate DSL wavefunctions~\cite{he2017signatures,hu2019dirac} is an interesting direction to pursue. 

\section{Discussion}
\label{sec:discussion}
This work uses the viewpoint that certain magnetically ordered states can be obtained upon condensing monopole operators which enter the low energy description of a Dirac spin liquid. We have argued that by using a Josephson junction geometry with two ordered states separated by a DSL, one can induce an emergent electric field (both DC and AC) in the DSL. Further, we have shown that such an AC emergent electric field can be measured optically as a sharp field-tunable peak in Raman scattering. Also, the induced electric field is accompanied by a measurable spin current across the junction that is proportional to the emergent electric field. This serves as an independent check that can be used to validate our first prediction. We have also highlighted other phenomena conceptually related to the monopole Josephson effect, namely long-range phase rigidity between puddles of ordered phases separated by Dirac spin liquids, ``mixed currents" across AFM-DSL-VBS junctions and spinon charge bound to lattice dislocations.

In general, an AFM -- X -- AFM junction, for some unknown phase X and generic details of the interface, will (at least in the short-junction limit) allow monopole tunneling analogous to that of the AFM -- DSL -- AFM junction. The observation of, for example, a spin current in a DC junction setup is therefore insufficient to claim that the unknown phase X is a DSL. However, two of the effects we propose to measure, namely the field-tunable Raman peak and the power-law dependence of the spin current as a function of junction size, require a conserved monopole current in region X (or equivalently require that the low-energy degrees of freedom of region X include an emergent $U(1)$ electric field). This requirement is satisfied when X is a DSL, but not in ordered phases such as valence bond solids where a spin singlet monopole creation operator acquires nonzero expectation value, and hence the monopole current (electric field) is not a conserved quantity. Therefore, measuring our proposed field-tunable sharp Raman peak in the region X in conjunction with a spin current across the interface, such that both the strength of the Raman peak and the spin current scale as a power law in the width of region X, will be strong evidence that X is a DSL.

We note that our predictions are backed up by writing a phenomenological monopole tunneling Hamiltonian that assumes that a DSL couples to a nearby ordered state chiefly through monopole tunneling terms, since monopoles are the most relevant operators in the DSL. We do not however attempt a full boundary conformal field theory calculation. This is because the current understanding of QED$_3$ as a CFT (even without boundaries) is still in its nascent stages, although there have been promising recent numerical developments~\cite{chester2016towards,albayrak2022bootstrapping,he2021conformal}. We also note that due to the Josephson effect, the quantum state of the DSL region differs from the ground state of QED$_3$. For example, when there is an AC electric field through the DSL, the DSL is in a nonequilibrium state. When there is a mixed spin-valley current, lattice translation symmetry gets broken. In such cases, whether the framework of DSL theory is still a valid description or not would depend on the strength of coupling between the different regions, size of the DSL region, and temperature. Determining this would again require a detailed boundary CFT calculation, and is beyond the scope of this work.

Our work presents an in-principle method to externally induce and measure emergent gauge field strengths in strongly coupled spin liquids in 2+1 dimensions. In general, one has more control over the degrees of freedom in an ordered state. So looking forward, attempting to probe operators in other spin liquids using more conventional ordered states is a promising direction. We note that Ref.~\cite{nakosai2019magnetic} theoretically considered tunneling of spinons between ferromagnets through a quantum spin ice in 3+1D, and is closely related to this idea.

A second interesting direction is in the context of recent developments in Rydberg atom arrays that take us one step closer to realizing a spin liquid in a lab~\cite{semeghini2021probing}. In these experiments, one can access projections of the microscopic wavefunction in a preferred basis. It will therefore be interesting to come up with signatures of long wavelength operators and non-equilibrium steady state features such as currents, but in the many-body wave function, such that they can be accessed in these experiments. 

\textit{Note added:} Just before the submission of this work, we became aware of another recent work~\cite{balentsBilayerDSL} considering monopole tunneling in Dirac spin liquids.

\begin{acknowledgments}
Some ideas and calculations on Raman scattering probes of spin liquids presented in Sec.~\ref{sec:raman} were inspired by the independent ongoing work by Mohammad Hafezi and collaborators (to be published). The authors are grateful to Mohammad Hafezi for useful discussions on this and other topics.  This work was supported by the National Science Foundation under Grant No. DMR-2037158, the U.S. Army Research Office under Contract No. W911NF1310172, and the Simons Foundation (V.G. and G.N.).  D.B. was supported by JQI-PFC-UMD. 
\end{acknowledgments}

\appendix

\section{Review of stability of DSL}
\label{app:stabilityreview}
If the DSL were to be a stable CFT, then it should contain no relevant (scaling dimension $\Delta>3$) symmetry allowed operators. In Table~\ref{tab:scaldim} , we summarize the scaling dimensions $\Delta$ of some of the important operators derived by references~\cite{dyer2013monopole,rantner2001electron,borokhov2003topological,dupuis2021anomalous} in the large $N_f$ limit. 

\begin{table}[h!]
    \centering
    \setlength{\extrarowheight}{.6ex}
    \begin{tabular}{|p{8em}|c|c|}\hline
        Operator & $\Delta_{N_f}$ & $\Delta_{N_f=4}$\\ \hline
        Monopoles & &\\ \hline
        $\vb{\hat{\Phi}}_{2\pi}$ & $0.2651 N_f - 0.0381 +\mathcal{O}(1/N_f)$ & 1.022\\ \hline
        $\vb{\hat{\Phi}}_{4\pi}$ & $0.6731 N_f - 0.1934+\mathcal{O}(1/N_f)$ & 2.499\\  \hline
        $\vb{\hat{\Phi}}_{6\pi}$ & $1.1864 N_f - 0.4211+\mathcal{O}(1/N_f)$ & 4.325\\ \hline \hline
        Fermion bilinears & & \\ \hline
        $\bar{\psi}\sigma^{\alpha}\tau^{\beta}\psi$, \footnotesize{($\alpha, \beta$ are not both $0$)} & $2-\frac{64}{3\pi^2 N_f}+\mathcal{O}(\frac{1}{N_f^2})$& 1.46\\ \hline
        $\bar{\psi}\psi$ & $2+\frac{128}{3\pi^2 N_f}+\mathcal{O}(\frac{1}{N_f^2})$ & 3.08\\ \hline \hline
        Conserved charges and currents & &\\ \hline
        $\hat{b}, \hat{e}_i, \hat{Q}\bqty{\sigma^\alpha \tau^\beta}$, $\hat{\vec{J}}\bqty{\sigma^\alpha \tau^\beta}$ & 2 &2\\ \hline
        
    \end{tabular}
   \caption{Scaling dimensions $\Delta_{N_f}$ of some important primary operators in QED$_3$, calculated in the large $N_f$ limit, compiled from~\cite{borokhov2003topological,dyer2013monopole,rantner2001electron,dupuis2021anomalous}. An operator with $\Delta>3$ is relevant in the RG sense.}
    \label{tab:scaldim}
\end{table}
 Ref.~\cite{song2020spinon} determined the symmetry properties of monopole operators for various lattices:
 \begin{enumerate}
     \item \underline{Bipartite lattices}: There is a symmetry allowed $2\pi$ monopole which is relevant according to the large $N_f$ analysis summarized in Table~\ref{tab:scaldim}. Hence a DSL cannot be a stable phase on bipartite lattices.
     \item \underline{Kagome lattice}: There is a symmetry allowed $4\pi$ monopole, which is likely relevant ($\Delta\approx 2.5$) according to the large $N_f$ calculation. 
     \item \underline{Triangular lattice}: $\vb{\hat{\Phi}}_{2\pi}$ and $\vb{\hat{\Phi}}_{4\pi}$ break translation symmetry and hence are symmetry-forbidden. (We will provide a more microscopic motivation for this fact in Sec.~\ref{sec:monopole}.) A $6\pi$ monopole operator is symmetry allowed, but is irrelevant ($\Delta=4.322$) according to the large $N_f$ calculation. This suggests that a DSL could indeed be a stable phase on the triangular lattice.
 \end{enumerate}
 So, in this work, whenever we refer to microscopic operators, we will assume a triangular lattice for concreteness. However, our general idea applies to any lattice which can realize a DSL as a stable phase.

\section{Microscopic expressions for field theory operators}
\label{sec:micro}
In this section, we construct microscopic operators corresponding to operators in the effective field theory. For a given field theory operator $\hat{\mathcal{O}}_{\text{tot}}\equiv \int d^2x \hat{\mathcal{O}}(x)$, we construct the microscopic operator
\begin{equation}
    \tot{\hat{\mathcal{O}}}=\sum_{\vec{n}}e^{i\vec{Q}.\vec{n}}\hat{\mathcal{O}}_{\vec{n}},
\end{equation}
where we have allowed for $\tot{\hat{\mathcal{O}}}$ to have momentum $\vec{Q}$ at the lattice scale. We use the following procedure~\cite{hermele2005algebraic,hermele2008properties,song2019unifying}:
\begin{enumerate}
    \item Find how $\tot{\hat{\mathcal{O}}}$ transforms under the microscopic symmetries. For operators that can be written in terms of fermionic partons, this can be done using information obtained by expanding around Dirac points~\cite{wen2004quantum}. We tabulate the transformation properties of the conserved charges of $G_{IR}$ in Table~\ref{tab:charges}, and conserved currents in Table~\ref{tab:currents}. 
    \item Construct operators order by order in size (maximum of weight, i.e., number of spins in the support of a local term, and diameter, i.e., extent of a local term) transforming identically as $\tot{\hat{\mathcal{O}}}$.
    \end{enumerate}
     We do this for the emergent electric and magnetic fields and spinon charge density in Sec.~\ref{sec:charges}. For monopole operators, this procedure is harder, and requires information at the lattice scale. Ref.~\cite{song2020spinon} did this using a Wannier center calculation. In Sec.~\ref{sec:monopole}, we will motivate their result using an independent approach involving the algebra of operators.
\begin{table}[ht]
    \centering
    \setlength{\extrarowheight}{1.2ex}
    \begin{tabular}{|c|c|c|c|c|c|c|}\hline
          Charge& $SO(3)$ & $\mathcal{T}$ & $T_1$ & $T_2$ & $C_6$ & $R_x$  \\ \hline
         $\tot{\hat{b}}$ & S & -1 & 1 & 1 & -1 & 1\\ \hline \hline
         $\hat{Q}\bqty{\sigma^i}$ & T & -1 & 1 & 1 & 1 & 1 \\ \hline \hline
         $\hat{Q}\bqty{\tau^1}$ & S & -1 & -1 & -1 & $\hat{Q}\bqty{\tau^2}$ & $\hat{Q}\bqty{\tau^3}$ \\ \hline
         $\hat{Q}\bqty{\tau^2}$ & S & -1 & 1 & -1 & $-\hat{Q}\bqty{\tau^3}$ & $-\hat{Q}\bqty{\tau^2}$ \\ \hline
         $\hat{Q}\bqty{\tau^3}$ & S & -1 & -1 & 1 & $-\hat{Q}\bqty{\tau^1}$ & $\hat{Q}\bqty{\tau^1}$ \\ \hline \hline
         $\hat{Q}\bqty{\sigma^i \tau^1}$ & T & 1 & -1 & -1 & $-\hat{Q}\bqty{ \sigma^i \tau^2}$ & $-\hat{Q}\bqty{\sigma^i \tau^3}$ \\ \hline
         $\hat{Q}\bqty{\sigma^i \tau^2}$ & T & 1 & 1 & -1 & $Q\bqty{ \sigma^i \tau^3}$ & $\hat{Q}\bqty{\sigma^i \tau^2}$ \\ \hline
         $\hat{Q}\bqty{\sigma^i \tau^3}$ & T & 1 & -1 & 1 & $\hat{Q}\bqty{\sigma^i \tau^1}$ & $-\hat{Q}\bqty{ \sigma^i \tau^1}$ \\ \hline 
    \end{tabular}
    \caption{Symmetry properties of conserved charges of $G_{IR}$. $SO(3)$ is spin-rotation (S and T stand for singlet and triplet under spin-rotation respectively.). $\mathcal{T}$ is time-reversal. $T_1$ and $T_2$ are lattice translations about $\vec{\mathrm{a}}_1$ and $\vec{\mathrm{a}}_2$ respectively. $C_6$ is rotation by $2\pi/6$ about a vertex. $R_x$ is reflection about $\vec{\mathrm{a}}_1$. }
    \label{tab:charges}
\end{table}

\bgroup
\def\arraystretch{1.8}
\begin{table}[ht]
    \centering
    \begin{tabular}{|c|c|c|c|c|c|c|}\hline
         Current & $SO(3)$ & $\mathcal{T}$ & $T_1$ & $T_2$ & $C_6$ & $R_x$  \\ \hline
         $\frac{\epsilon^{ij}}{2\pi}\hat{e}^j$  & S & 1 & 1 & 1 & $-V$ & $-V$\\ \hline \hline
         $\hat{\vec{J}}\left[\sigma^i\right]$ & T & 1 & 1 & 1 & $V$ & $V$  \\ \hline \hline
         $\hat{\vec{J}}\left[\tau^1\right]$ & S & 1 & -1 & -1 & $V$ as $\hat{\vec{J}}\left[\tau^2\right]$ & $V$ as $\hat{\vec{J}}\left[\tau^3\right]$ \\ \hline
         $\hat{\vec{J}}\left[\tau^2\right]$ & S & 1 & 1 & -1 & $V$ as $-\hat{\vec{J}}\left[\tau^3\right]$ & $V$ as $-\hat{\vec{J}}\left[\tau^2\right]$ \\ \hline
         $\hat{\vec{J}}\left[\tau^3\right]$ & S & 1 & -1 & 1 & $V$ as $-\hat{\vec{J}}\left[\tau^1\right]$ & $V$ as $\hat{\vec{J}}\left[\tau^1\right]$ \\ \hline \hline
         $\hat{\vec{J}}\left[\sigma^i \tau^1\right]$ & T & -1 & -1 & -1 & $V$ as $-\hat{\vec{J}}\left[\sigma^i \tau^2\right]$ & $V$ as $-\hat{\vec{J}}\left[\sigma^i \tau^3\right]$ \\ \hline
         $\hat{\vec{J}}\left[\sigma^i \tau^2\right]$ & T & -1 & 1 & -1 & $V$ as $\hat{\vec{J}}\left[\sigma^i \tau^3\right]$ & $V$ as $\hat{\vec{J}}\left[\sigma^i \tau^2\right]$ \\ \hline
         $\hat{\vec{J}}\left[\sigma^i \tau^3\right]$ & T & -1 & 1 & -1 & $V$ as $\hat{\vec{J}}\left[\sigma^i \tau^1\right]$ & $V$ as $-\hat{\vec{J}}\left[\sigma^i \tau^1\right]$ \\ \hline
         
    \end{tabular}
    \caption{Symmetry properties of conserved currents of $U(1)_{\text{top}}$ and $SO(6)$. Notation: ``$V$'' means ``transforms as a vector'', ``$-V$'' means transforms as a vector except for a factor of $-1$. ``$V$ as $\hat{\vec{J}}\bqty{\sigma^i \tau^j}$'' means the current's spatial indices are transformed as a vector while the $SO(6)$ indices are rotated to $\sigma^i \tau^j$, possibly with an overall sign. }
    \label{tab:currents}
\end{table}
\egroup
\subsection{Emergent electric and magnetic field}
\label{sec:charges}
The generator of $U(1)_{\text{top}}$ is the total emergent magnetic flux $\tot{\hat{b}}\equiv \frac{1}{2\pi}\int d^2 x \hat{b}(x)$. Because $\tot{\hat{b}}$ is odd under time-reversal (see Table~\ref{tab:charges}), and singlet under spin rotation, the lowest weight term is a 3-spin spin chirality: $\downnchi \equiv \chiof{\vec{n}}{\vec{n}+(1,-1)}{\vec{n}+(1,0)}$ and $\upnchi \equiv \chiof{\vec{n}}{\vec{n}+(1,0)}{\vec{n}+(0,1)}$. Here, we have used the notation $(n_1,n_2)\equiv n_1 \vec{\mathrm{a}}_1 + n_2 \vec{\mathrm{a}}_2$. If we only keep (1) elementary triangles $\pqty{\mysymb{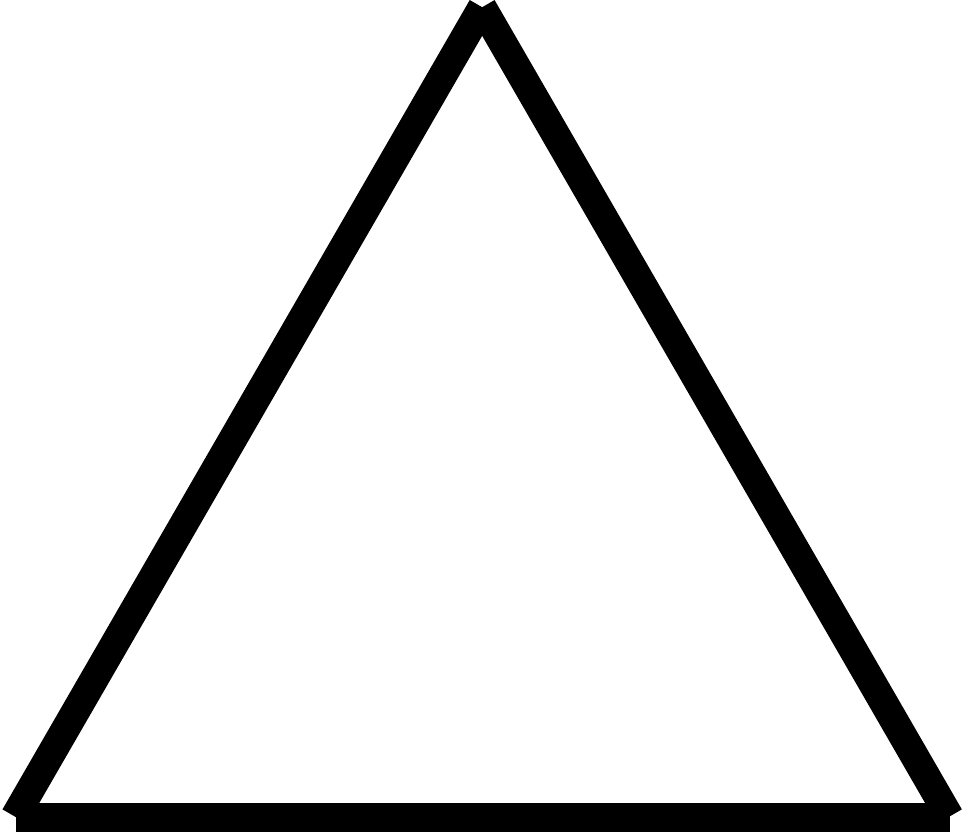}{2.2}}$ and (2) triangles whose 2 edges are nearest-neighbour $\pqty{\mysymb{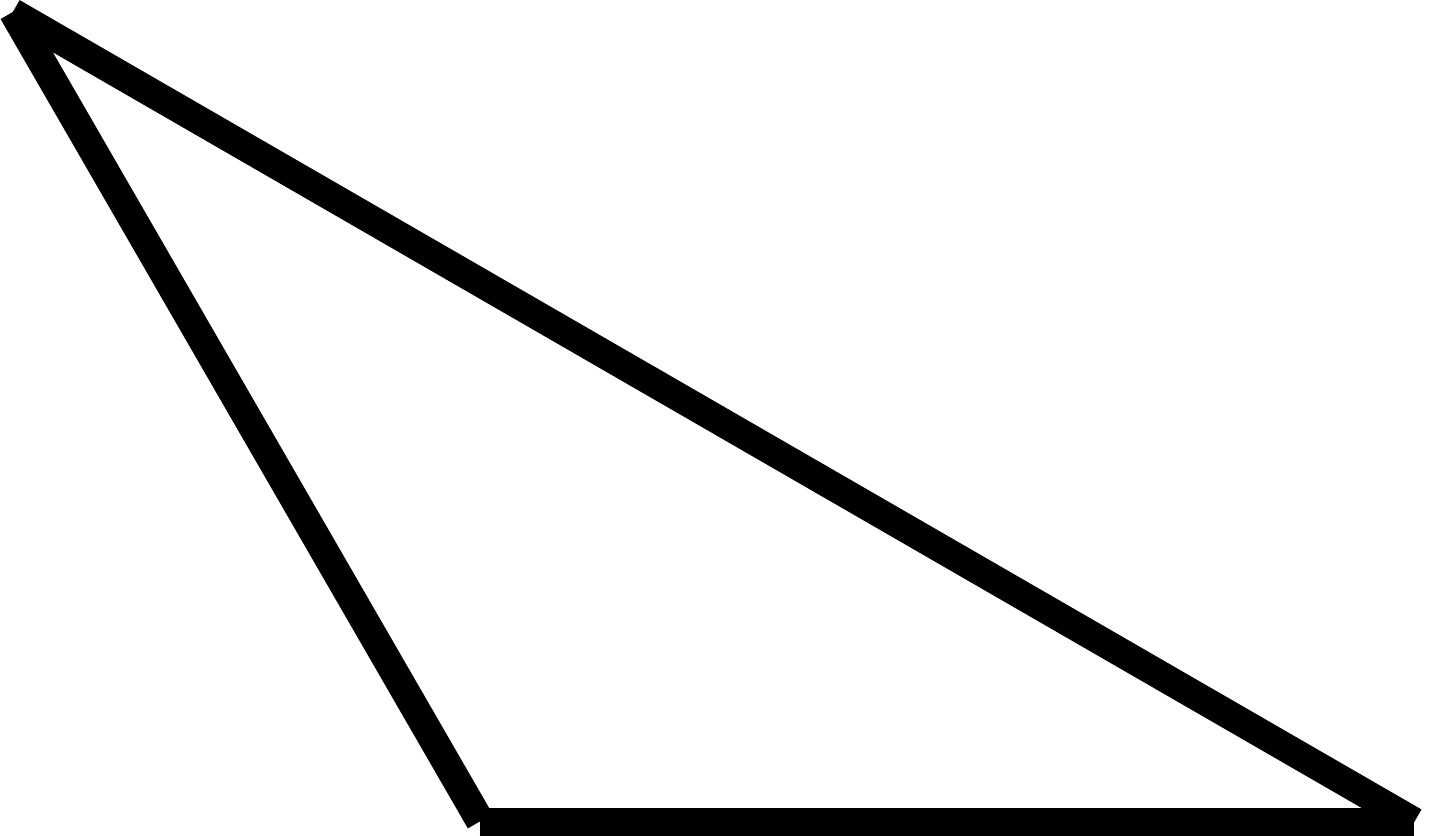}{2.2}}$, then the only term consistent with symmetries is 
\begin{equation}
\label{eq:bexp}
    \tot{\hat{b}}=\sum_{\vec{n}} \pqty{\downnchi - \upnchi} + \ldots
\end{equation}.

Since the total emergent magnetic flux is the $U(1)_{\text{top}}$ charge density, it follows from Faraday's law that the emergent electric field $\hat{\vec{e}}$ is the $U(1)_{\text{top}}$ conserved current rotated by $90^\circ$.

If we consider all operators for $\hat{\vec{e}}$ made of terms with 2 spins (both nearest neighbor and next-nearest-neighbor), then (in the notation: $\hat{\vec{e}}\equiv \hat{e}_1 \vec{\mathrm{a}}_1 + \hat{e}_2 \vec{\mathrm{a}}_2$),
\begin{equation}
\label{eq:el1}
    \hat{e}_i=\alpha_1 (\hat{e}_i)^{(1)} + \alpha_2 (\hat{e}_i)^{(2)} + \ldots \quad \text{for }i=1,2,
\end{equation}
where 
\begin{align}
\label{eq:el2}
    (\hat{e}_1)^{(1)}=\sum_{\vec{n}} \mysymb{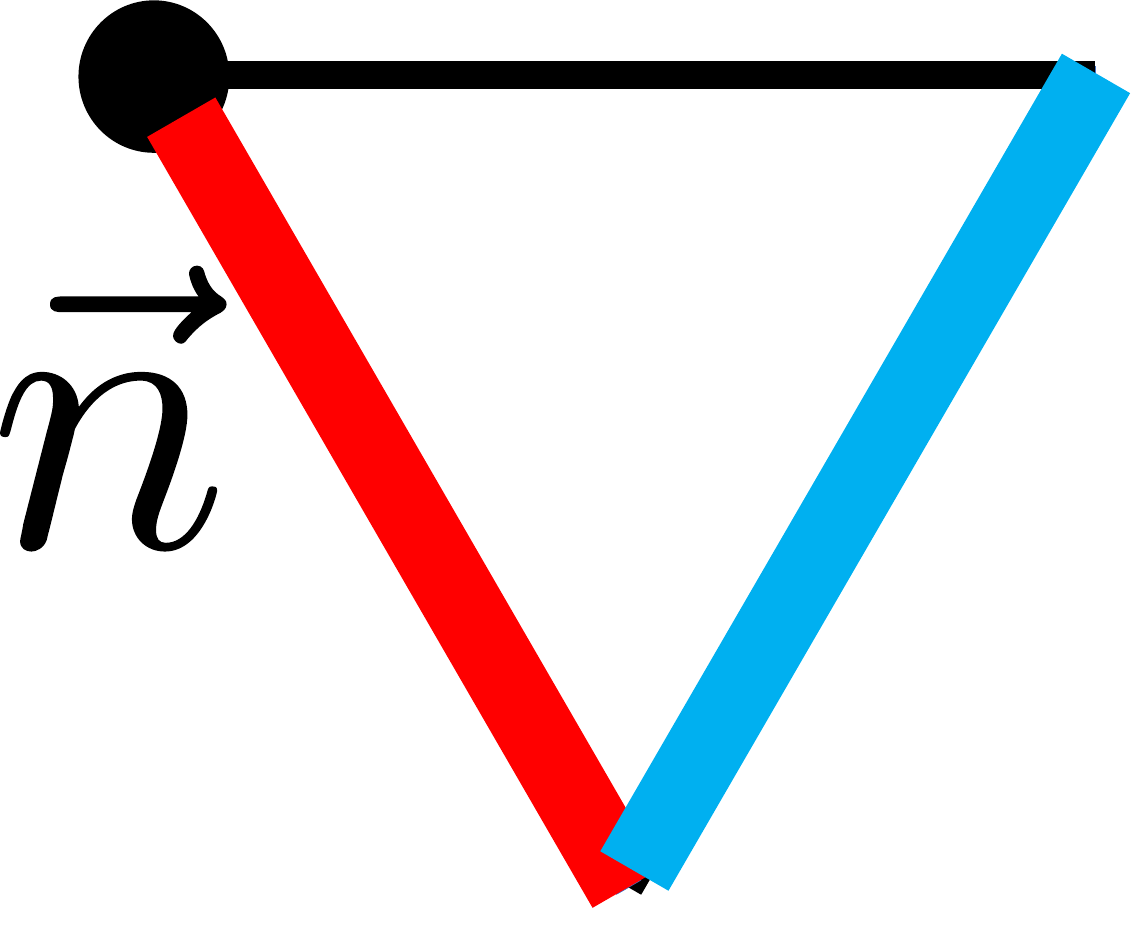}{4.0}  &, \quad (\hat{e}_2)^{(1)}=\sum_{\vec{n}} \mysymb{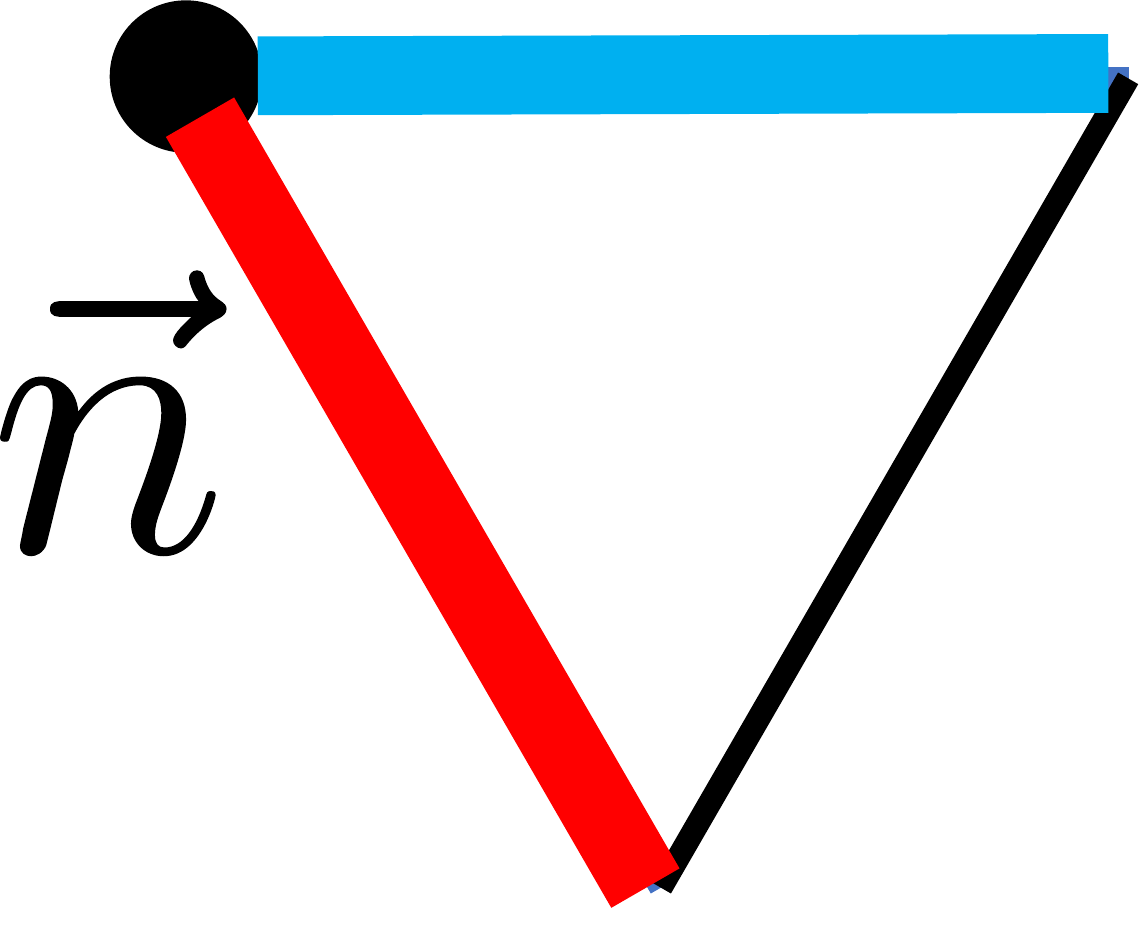}{4.0},\\
    (\hat{e}_1)^{(2)}=\sum_{\vec{n}} \mysymb{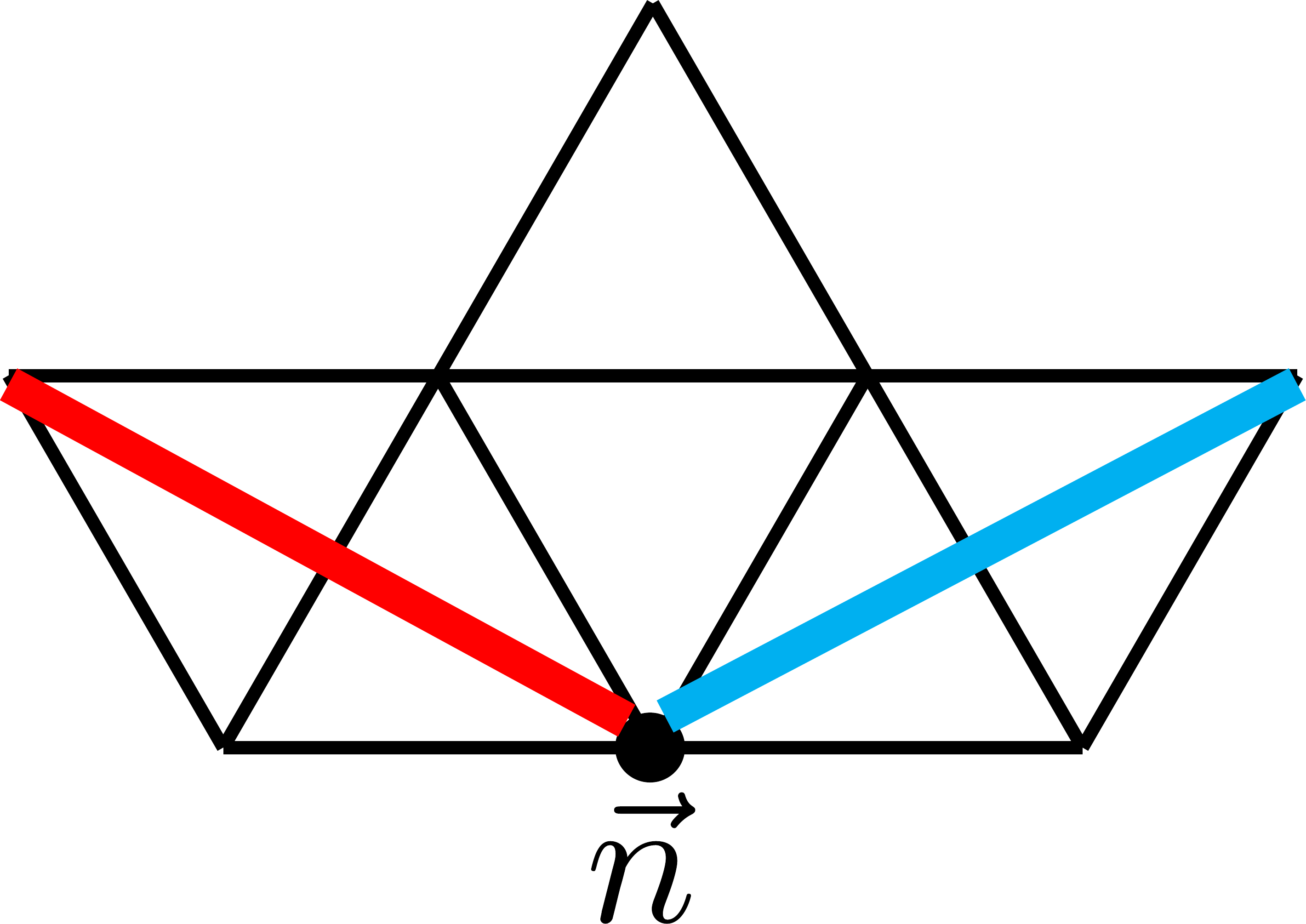}{8.0} &, \quad (\hat{e}_2)^{(2)}=\sum_{\vec{n}} \mysymb{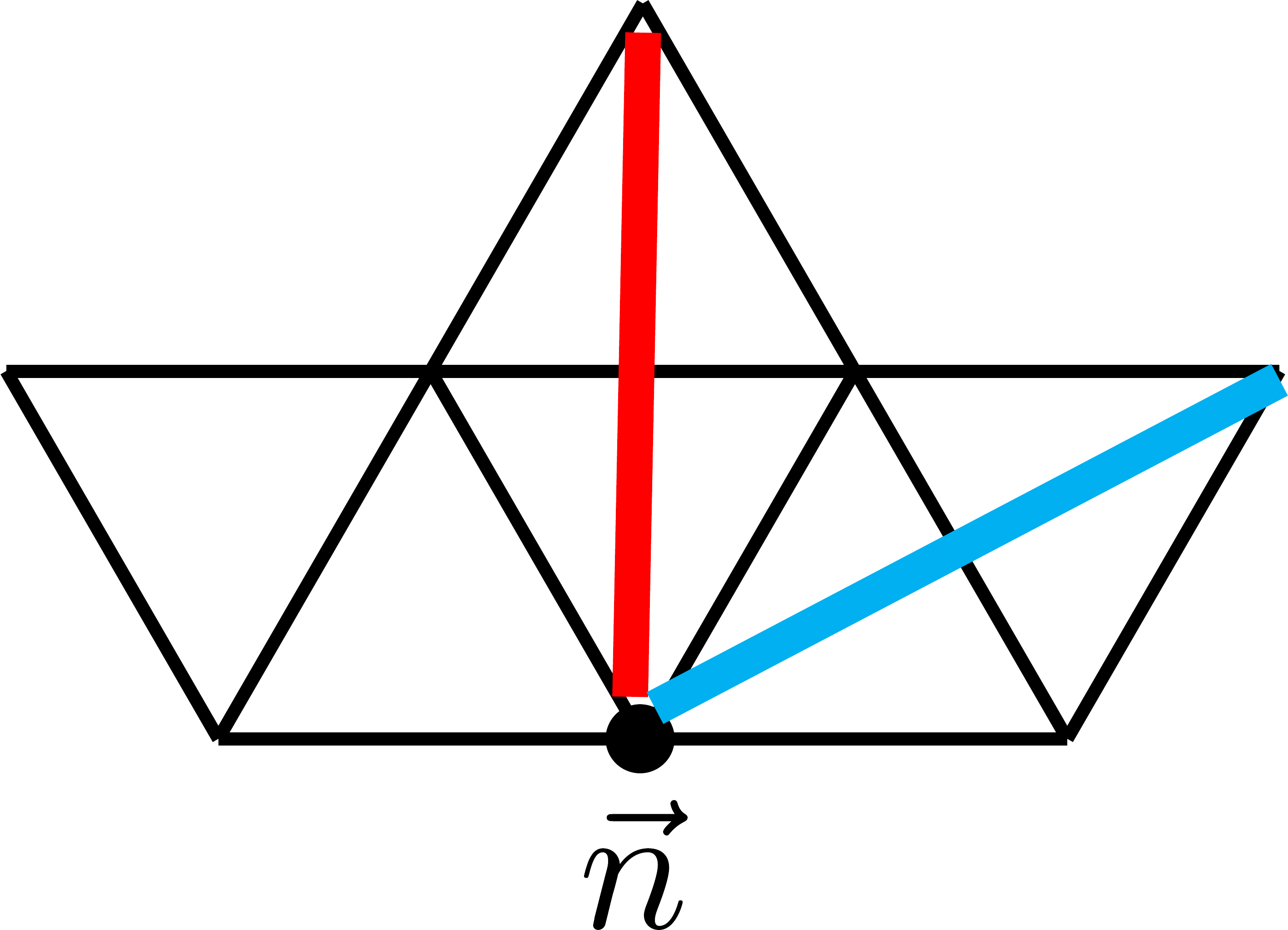}{8.0},
\end{align}
with the notation
\begin{equation}
\begin{aligned}
    \mysymb{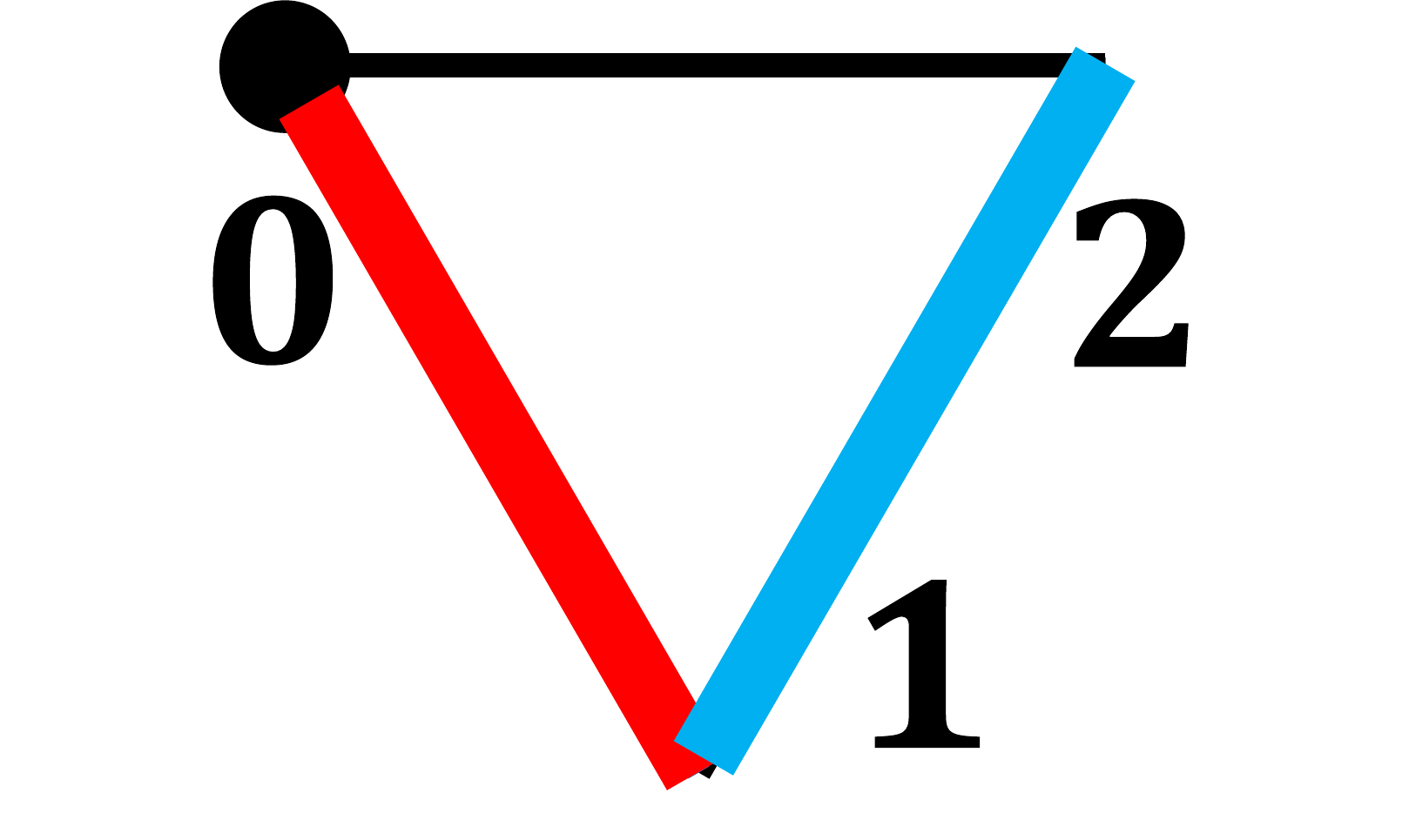}{4.0}&\equiv \dotof{1}{2}-\dotof{0}{1} \text{ and }\\
    \mysymb{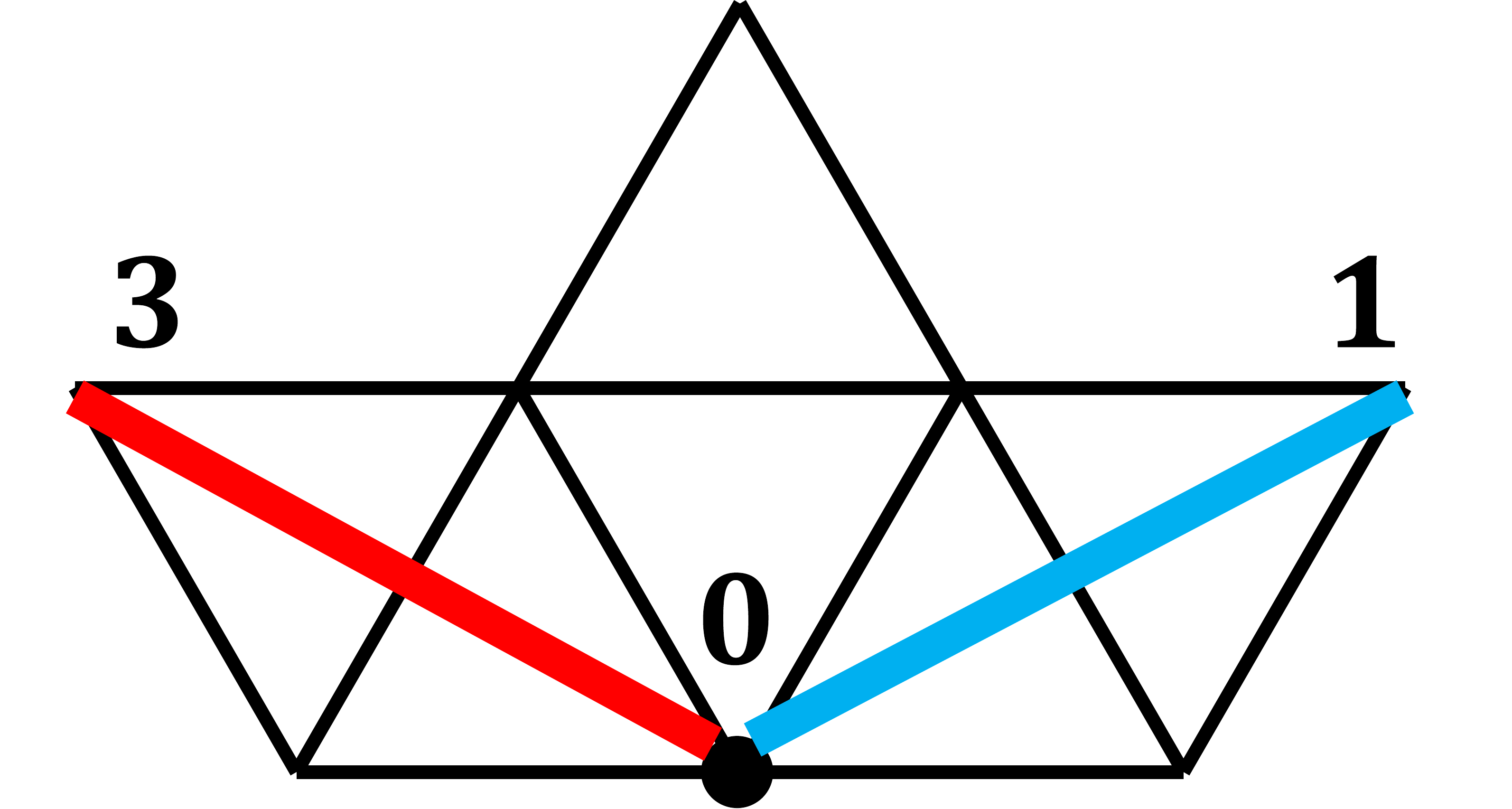}{9.0}&\equiv \dotof{0}{1}-\dotof{0}{3},
    \end{aligned}
\end{equation}
Note that $\hat{e}_1$ and $\hat{e}_2$ are not orthogonal. $\hat{e}_x$ and $\hat{e}_y$ are related to $\hat{e}_1$ and $\hat{e}_2$ as $\hat{e}_x=\hat{e}_1$ and $\hat{e}_y=\frac{1}{\sqrt{3}}(-\hat{e}_1 +2 \hat{e}_2)$.

From the above expression for electric field, one can compute the local divergence of the electric field, which is proportional to the spinon charge by Gauss's law:
\begin{equation}
    \hat{q}=\frac{1}{g^2}\mathrm{div} \hat{\vec{e}}
\end{equation}
We see that only $\hat{\vec{e}}^{(2)}$ contributes to $\hat{q}$, and not $\hat{\vec{e}}^{(1)}$. Therefore, $\hat{\vec{e}}^{(1)}$ is the transverse electric field and $\hat{\vec{e}}^{(2)}$ is the longitudinal electric field. $\hat{q}$ at site $i$ is given by
\begin{equation}
\label{eq:expcharge}
\begin{split}
    \hat{q}_i =& \mysymb{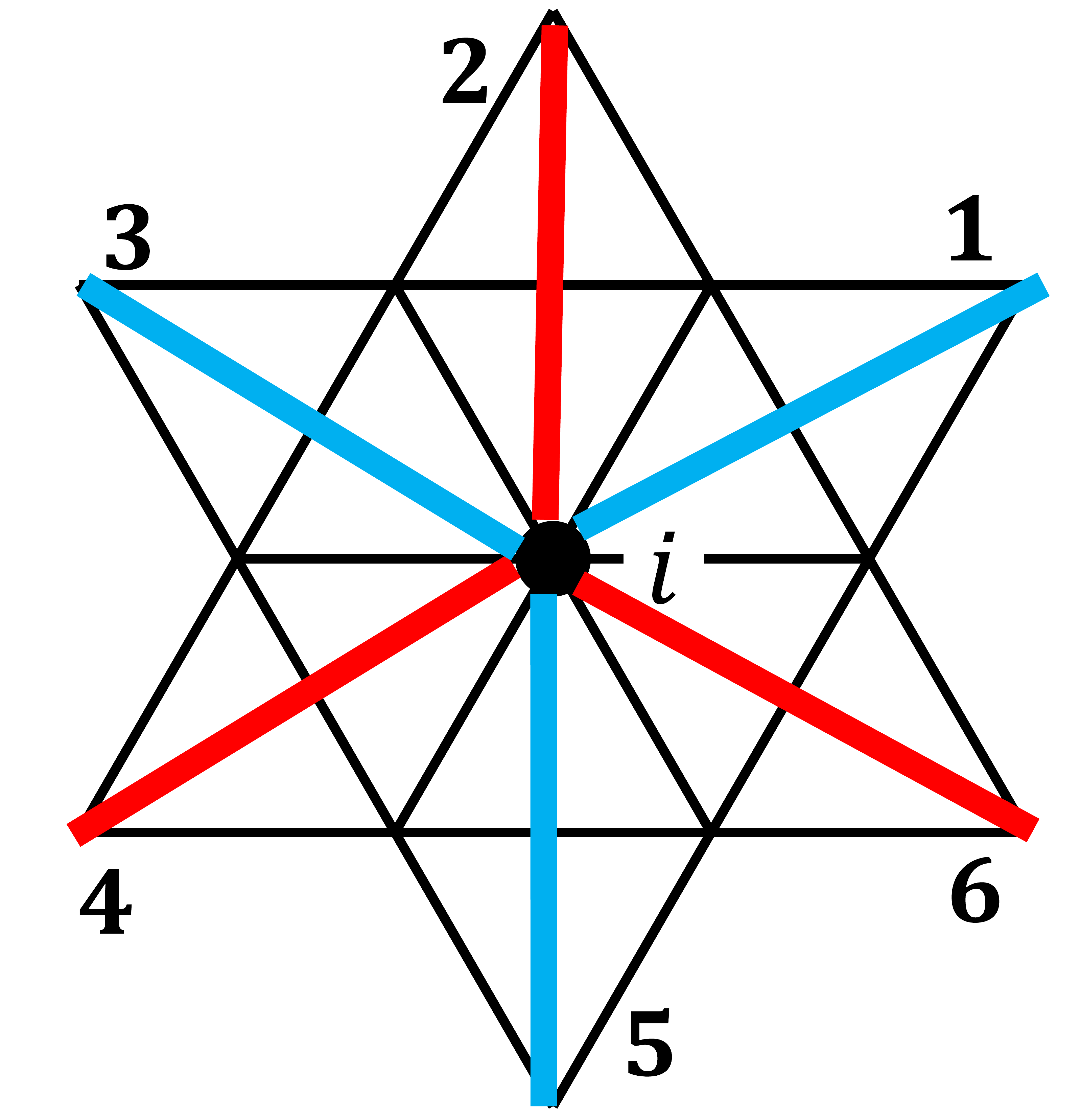}{16.0}+\ldots \equiv \left(\dotof{i}{1}-\dotof{i}{2}+\right.\\
    &\left.+\dotof{i}{3}-\dotof{i}{4}+\dotof{i}{5}-\dotof{i}{6}\right)+\ldots
\end{split}.
\end{equation}
By construction,  we see that the sum of spinon charge enclosed in a region $D$ is an operator with support localized to the boundary of $D$. This is consistent with Gauss's law. It also satisfies $\expval{\hat{q}_i}=0$ by symmetry.
\subsection{Monopole operators from commutation relations}
\label{sec:monopole}
The symmetry properties of monopole operators were calculated in~\cite{song2019unifying,song2020spinon} and reported in Table 2 of Ref.~\cite{song2020spinon}. The $SO(6)$ contribution was calculated using Step 1 (beginning of Appendix \ref{sec:micro}). The $U(1)_{\text{top}}$ contribution was calculated using a Wannier center calculation of the free fermion bands for the mean field ansatz. Here, we attempt an alternative approach to calculate the $U(1)_{\text{top}}$ contribution. While our calculation involves an uncontrolled approximation, it provides an independent motivation for the result in Ref.~\cite{song2019unifying}.  

The principle behind our approach is that the algebra of $G_{IR}$ has to be obeyed down to the microscopic level because we are dealing with operators of the form $\hat{\mathcal{O}}_{\text{tot}}$ here, which have the longest possible wavelength (allowed by their symmetry properties):
    \begin{equation}
    \label{eq:commQ}
    \comm{\tot{\hat{Q}}^{ab}}{\tot{\hat{Q}}^{cd}}=i\pqty{\delta_{bc}\tot{\hat{Q}}^{da}-\delta_{ac}\tot{\hat{Q}}^{db}+\delta_{ad}\tot{\hat{Q}}^{cb}-\delta_{bd}\tot{\hat{Q}}^{ca}},
    \end{equation}
    \begin{equation}
    \label{eq:commb}
    \comm{\tot{\hat{b}}}{\tot{\hat{Q}}^{ab}}=0.
\end{equation}
Here $\{\hat{Q}^{ab}\}$ (antisymmetric in $a,b$ with $a,b$ running from 1 to 4) are the 15 generators of $SO(6)$ and $\tot{\hat{b}}$ is the generator of $U(1)_{\text{top}}$ (see Appendix~\ref{app:notation} for the notation). 

Next, $2\pi$ monopole operators that are charged under $G_{IR}$ have to obey the algebra
\begin{align}
    \label{eq:commMb}
     \comm{\tot{\hat{b}}}{\tot{(\hat{\Phi}^\dagger_j)}}&= \tot{(\hat{\Phi}^{\dagger}_j)},    \\
    \label{eq:commMQ}
    \comm{\tot{\hat{Q}}^{bc}}{\tot{(\hat{\Phi}^\dagger_j)}}&=\sum_{i=1}^6 \tot{(\hat{\Phi}^\dagger_i)}\pqty{T^{bc}}_{ij}.
\end{align}
where, $T^{bc}$, a matrix of c-numbers, is the generator of $\tot{\hat{Q}}^{bc}$ acting on $\mathbb{C}^6$, and the matrix elements are given by
\begin{equation}
\label{eq:Tdef}
    \pqty{T^{bc}}_{ij}=-i(\delta_{cj}\delta_{ib}-\delta_{bj}\delta_{ic}).
\end{equation}
This suggests a general procedure:
\begin{enumerate}
    \item Suppose $\tot{\hat{\mathcal{O}}}=\sum_{\vec{n}}e^{i\vec{Q}.\vec{n}}\hat{\mathcal{O}}_{\vec{n}}$ If we now expand $\hat{\mathcal{O}}_{\vec{n}}$ in operators of increasing ``size" $s$, 
    \begin{equation}
        \hat{\mathcal{O}}_{\vec{n}}=\sum_{s=1}^{\infty} C_s \pqty{\hat{\mathcal{O}}_{\vec{n}}}_{s}.
    \end{equation}
    Here, each $\pqty{\hat{\mathcal{O}}_{\vec{n}}}_{s}$ is chosen to respect the symmetry properties obtained just from the low energy theory.
    \item Demand Eq.~\eqref{eq:commQ}-\eqref{eq:commb}, \eqref{eq:commMb}-\eqref{eq:commMQ} order by order in size $s$, and obtain constraints on $C_s$. 
\end{enumerate}
Here we will only perform this calculation at the lowest order in size by enforcing Eq.~\eqref{eq:commMb} up to a proportionality constant. 
\begin{equation}
\label{eq:commMbprop}
    \comm{\tot{\hat{b}}}{\hat{\va{\Phi}}^\dagger_{\text{tot}}}=\mathcal{K}\hat{\va{\Phi}}^\dagger_{\text{tot}},
\end{equation}
where $\mathcal{K}$ is a positive constant. Let us assume that the monopole inserting $2\pi$ flux is ``simpler'', i.e., has a lower leading operator size than the one inserting $4\pi$ or $6\pi$ flux. Then we ask what the ``simplest" spin triplet monopole operator is. We start with operators with size 1. 
\begin{equation}
\label{Phiguess}
    \hat{\va{\Phi}}^\dagger_{\text{tot}} = \sum_{\vec{n}} e^{i\vec{Q}\cdot\vec{n}} \Sof{\vec{n}} + \ldots
\end{equation}
From compatibility of translation symmetry with rotation symmetry, $\vec{Q}$ is either $(0,0)$ or $\pm(2\pi/3, -2\pi/3)$~\cite{song2019unifying}. Using identity Eq.~\eqref{eq:id1}, we evaluate the commutator in Eq.~\eqref{eq:commMbprop} using the lowest order expression for $\tot{\hat{b}}^{(1)}$ in Eq.~\eqref{eq:bexp}. Each single-spin term in $\hat{\va{\Phi}}^\dagger_{\text{top}}$ fails to commute with exactly 6 triangles in $\tot{\hat{b}}^{(1)}$. After evaluating each of these commutators, we get 
\begin{eqnarray}
\label{eq:monopole1}
    &&\comm{\tot{\hat{b}}^{(1)}}{\va{\Phi}^\dagger_{\text{tot}}}=\nonumber \\&&-i \sum_{\vec{n}}e^{i\vec{Q}\cdot\vec{n}} \Sof{\vec{n}}\left\{(e^{-iQ_1}  -e^{-iQ_2})(\dotof{\vec{n}_6}{\vec{n}_1})\right.\nonumber \\&& \left.+(e^{i(Q_1-Q_2)} -e^{iQ_2})(\dotof{\vec{n}_1}{\vec{n}_2})\right.\nonumber \\&&+(e^{i(Q_1-Q_2)} -e^{iQ_1})(\dotof{\vec{n}_2}{\vec{n}_3})\nonumber \\&&+(e^{iQ_2} -e^{iQ_1})(\dotof{\vec{n}_3}{\vec{n}_4}) \nonumber \\&&+(e^{iQ_2} -e^{i(Q_2-Q_1)})(\dotof{\vec{n}_4}{\vec{n}_5})\nonumber\\&&\left.(e^{-iQ_1} -e^{i(Q_2-Q_1)})(\dotof{\vec{n}_5}{\vec{n}_6})\right\}+\ldots,
\end{eqnarray}
where $\vec{n}_1\equiv \vec{n}+(0,-1)$, $\vec{n}_2\equiv \vec{n}+(1,-1)$, $\vec{n}_3\equiv \vec{n}+(1,0)$, $\vec{n}_4\equiv \vec{n}+(0,1)$, $\vec{n}_5\equiv \vec{n}+(-1,1)$ and $\vec{n}_6\equiv \vec{n}+(0,-1)$.
From this, it is clear that $(Q_1,Q_2)=(0,0)$ will give $0$ as the commutator. Therefore if the monopole operator is to have single spin terms as its leading order term, then $\vec{Q}=\pm(2\pi/3, -2\pi/3)$. For $\mathcal{K}$ in Eq.~\eqref{eq:commMbprop} to be positive, we choose $\vec{Q}=(2\pi/3, -2\pi/3)$. Here, we have made use of the fact that the DSL is a ground-state of an antiferromagnetic Heisenberg-like Hamiltonian where $\expval{\dotof{i}{j}}<0$ for nearest neighbors $i$ and $j$.  So, $(Q_1,Q_2)=(2\pi/3,-2\pi/3)$. With this choice, the Eq.~(\ref{eq:monopole1}) becomes
\begin{equation}
\label{eq:monopole2}
    \comm{\hat{b}_{\text{tot}}^{(1)}}{\va{\Phi}^\dagger_{\text{tot}}}=-\sqrt{3}\sum_{\vec{n}}e^{i\vec{Q}\cdot\vec{n}} \Sof{\vec{n}}\pqty{\mysymb{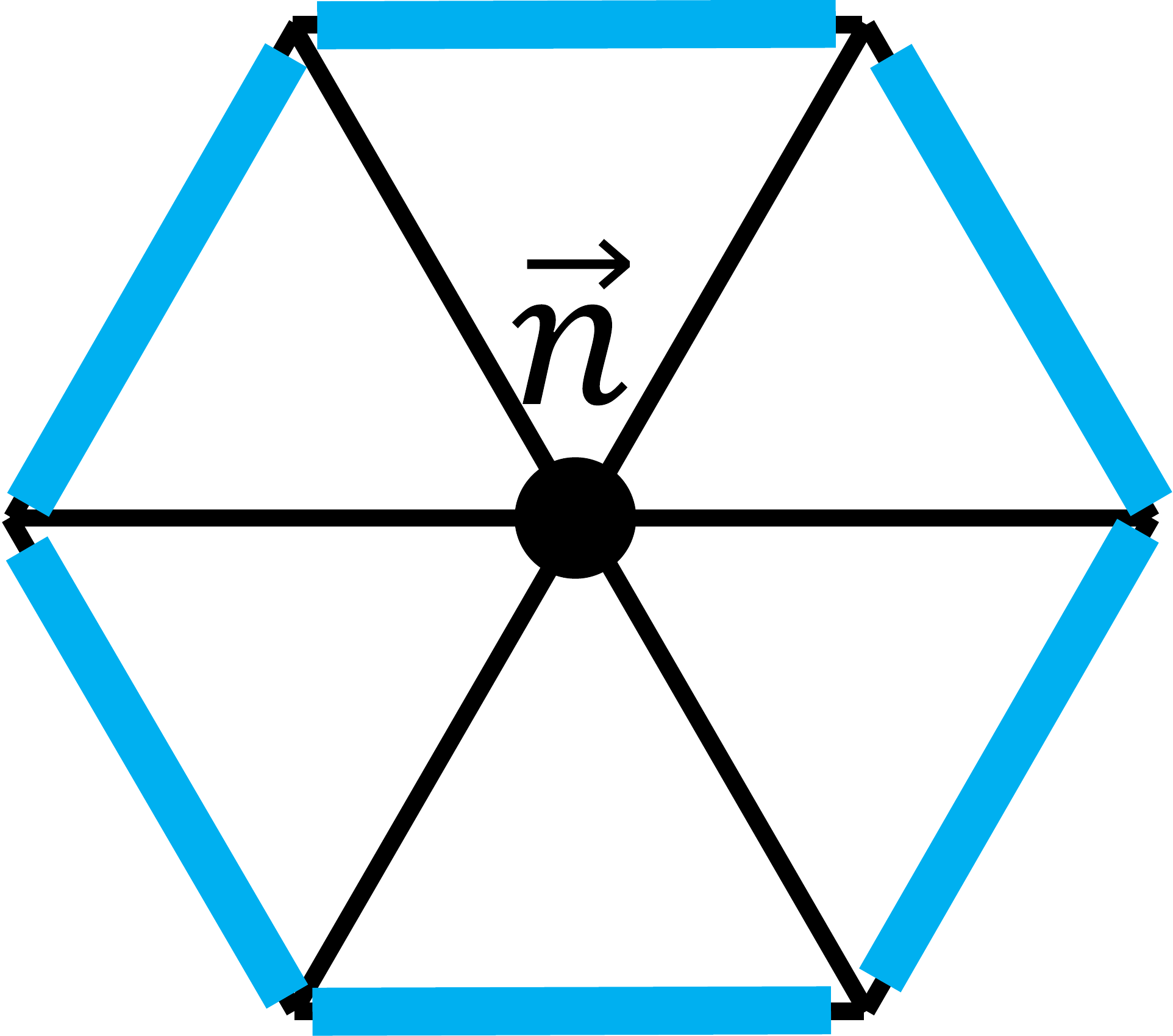}{7.2}}+\ldots,
\end{equation}
where we use the notation,
\begin{equation}
\begin{aligned}
    \mysymb{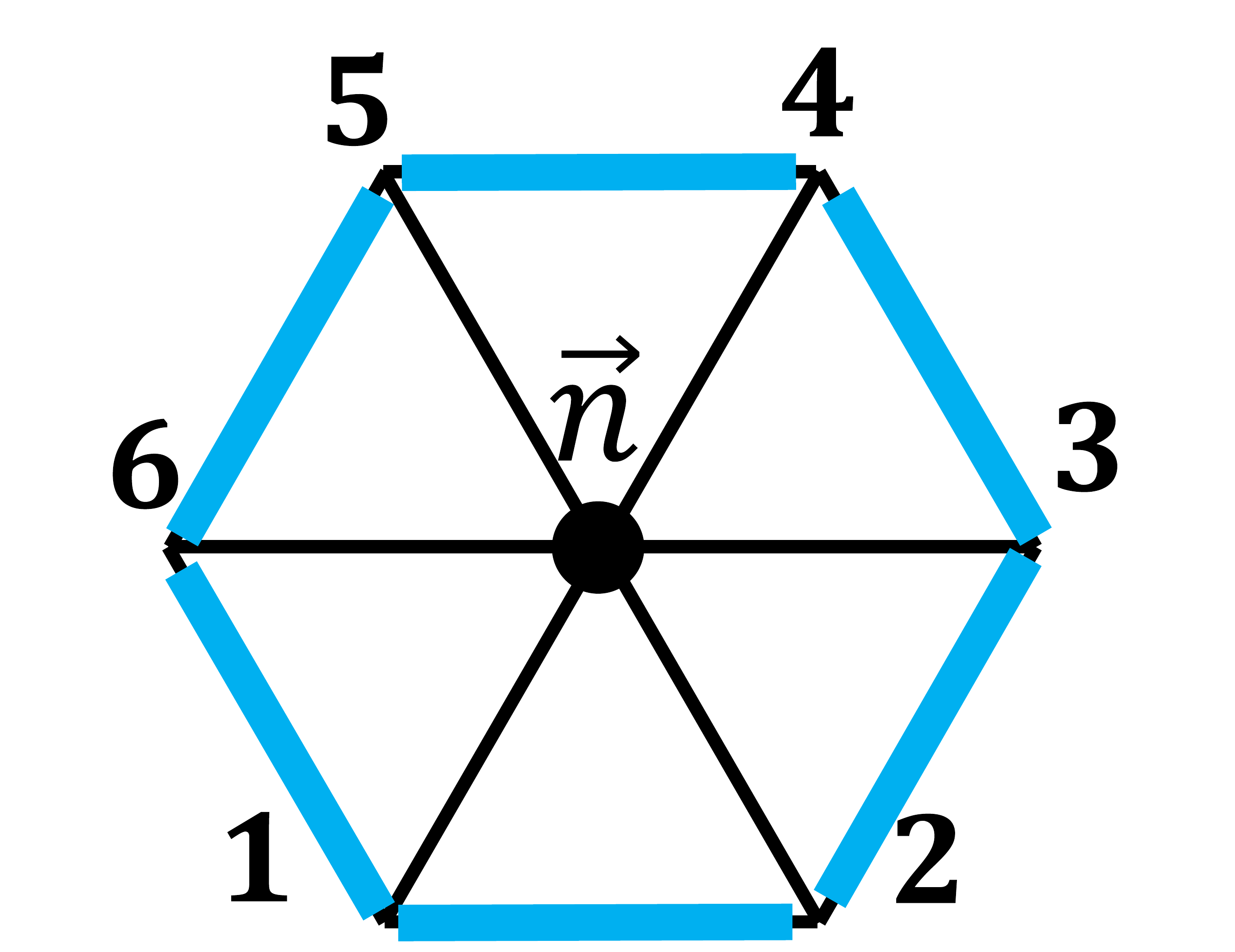}{8.4}&\equiv\dotof{1}{2}+\dotof{2}{3}+\dotof{3}{4}+\dotof{4}{5}\\ &+\dotof{5}{6}+\dotof{6}{1} \equiv \hat{\hexagon}_{\vec{n}}.
    \end{aligned}
\end{equation}
 To our guess for $\va{\hat{\Phi}}^\dagger_{\text{top}}$, we now add the RHS obtained above.
\begin{equation}
    \hat{\va{\Phi}}^\dagger_{\text{tot}} = \beta_1 \hat{\va{\Phi}}^{\dagger (1)} +\beta_2 \hat{\va{\Phi}}^{\dagger (2)}+ \ldots,
\end{equation}
where 
\begin{equation}
    \hat{\va{\Phi}}^{\dagger (1)}=\sum_{\vec{n}} e^{i\vec{Q}\cdot\vec{n}} \Sof{\vec{n}},\ \text{ and }\hat{\va{\Phi}}^{\dagger (2)}=\sum_{\vec{n}} e^{i\vec{Q}\cdot\vec{n}} \Sof{\vec{n}} \hat{\hexagon}_{\vec{n}}.
\end{equation}
We can use the commutators Eq.~ (\ref{eq:id1},\ref{eq:id8},\ref{eq:id9},\ref{eq:id10}) to get
\begin{equation}
\begin{aligned}
    \comm{\tot{\hat{b}}^{(1)}}{ \hat{\va{\Phi}}^{\dagger (1)}}&=-\sqrt{3} \hat{\va{\Phi}}^{\dagger (2)}  \text{ and }\\\comm{\tot{\hat{b}}^{(1)}}{ \hat{\va{\Phi}}^{\dagger (2)}}&=-\sqrt{3}\pqty{\frac{3}{4}\hat{\va{\Phi}}^{\dagger (1)} - \hat{\va{\Phi}}^{\dagger (2)} + \ldots}.
    \end{aligned}
\end{equation}
Using the above equation, truncating at terms supported on at most elementary triangles, we get
\begin{equation}
\label{eq:tripletexp}
    \hat{\va{\Phi}}^\dagger_{\text{tot}} = \sum_{\vec{n}}e^{i\vec{Q}.\vec{n}} \pqty{\Sof{\vec{n}}-\frac{4}{3}\Sof{\vec{n}} \hat{\hexagon}_{\vec{n}}+\ldots},
\end{equation}
where $\vec{Q}=(2\pi/3,-2\pi/3)$.
\subsubsection{Spin singlet monopoles}
Having determined the momentum of the spin-triplet $2\pi$-monopoles, the momenta of spin-singlet monopoles can be fixed by the low energy theory since the embedding of the space-group symmetries into $SO(3)_{\text{valley}}$ can be computed purely from low energy information. Doing so results in Table 2 of Ref.~\cite{song2019unifying}. Here, we will write microscopic expressions for them.

The spin singlet monopoles are time-reversal even. Here, we will only keep the lowest weight terms that are dot products of neighbouring spins:
\begin{equation}
\label{eq:singletexp}
    \hat{\Phi}^\dagger_i = v_1 \hat{\Phi}^{\dagger(1)}_i + v_2 \hat{\Phi}^{\dagger(2)}_i \quad \text{for }i\in\{1,2,3\},\text{ where }
\end{equation}
\begin{equation}
\begin{aligned}
\label{eq:singletm1}
    \hat{\Phi}^{\dagger(1)}_1&=e^{-i\frac{\pi}{3}}\sum_{\vec{n}}e^{i(-\frac{\pi}{3},\frac{\pi}{3})\cdot\vec{n}}   \pqty{ \mysymb{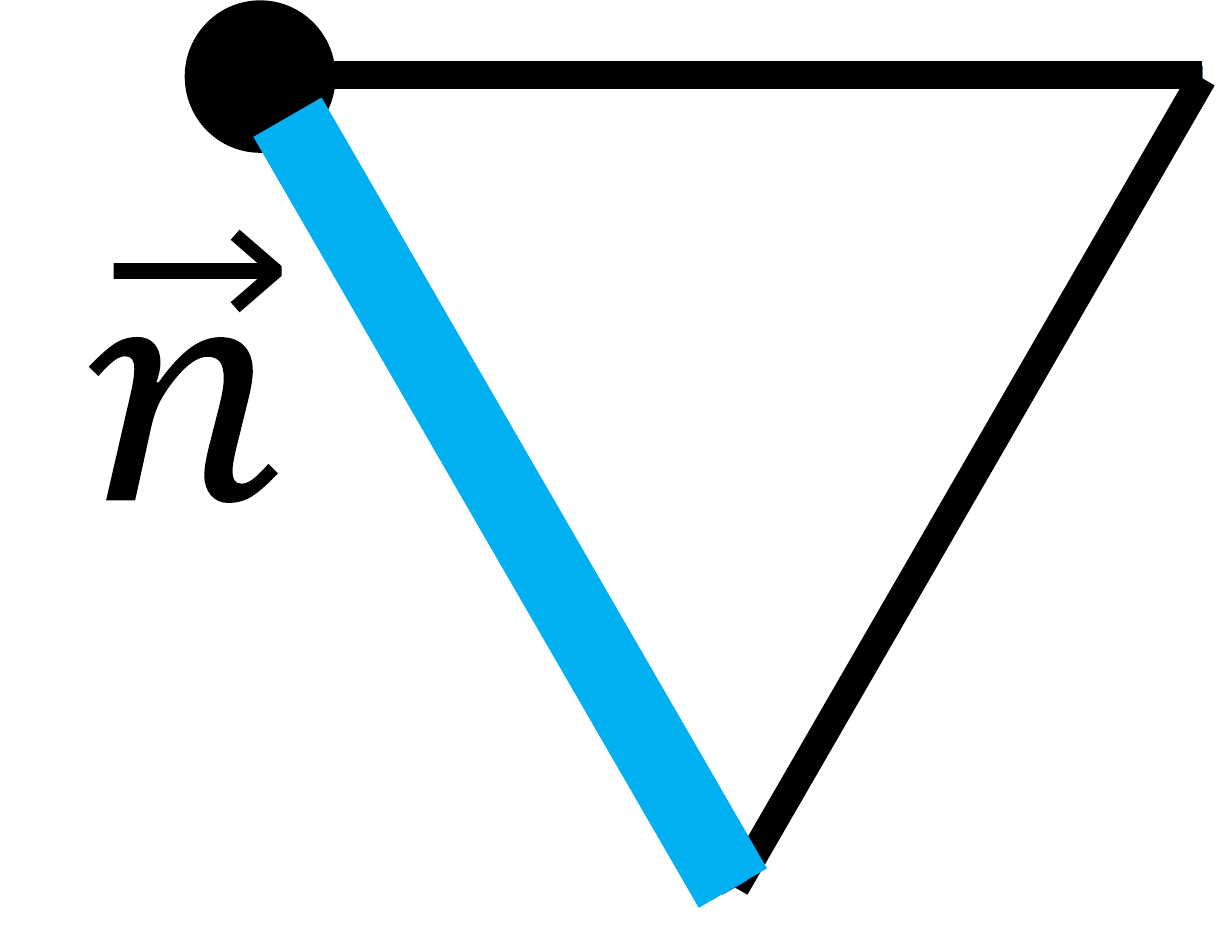}{4.0}}, \\
    \hat{\Phi}^{\dagger(1)}_2&= e^{i\frac{\pi}{3}}\sum_{\vec{n}}e^{i(\frac{2\pi}{3},\frac{\pi}{3})\cdot\vec{n}} \pqty{ \mysymb{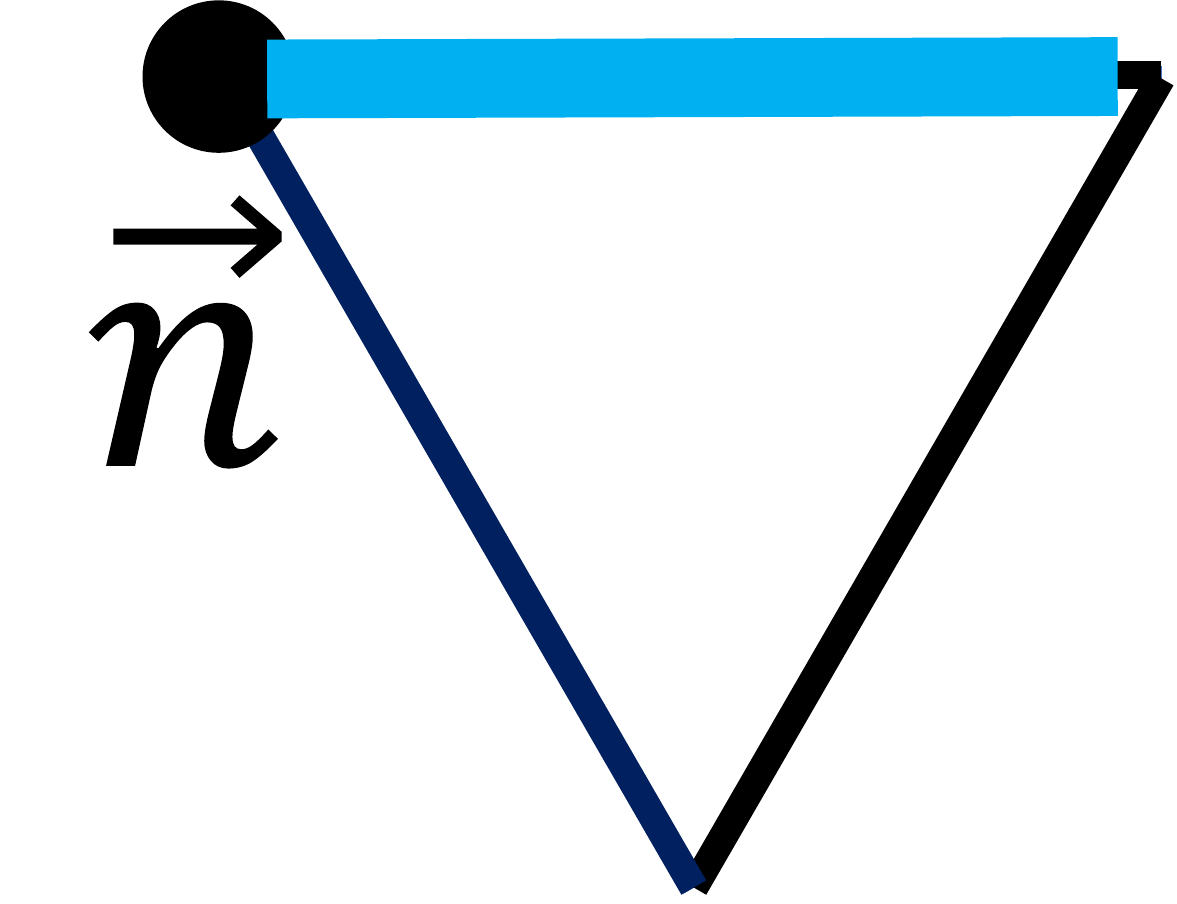}{4.0}},\\
     \hat{\Phi}^{\dagger(1)}_3&=-\sum_{\vec{n}}e^{i(-\frac{\pi}{3},-\frac{2\pi}{3})\cdot\vec{n}}\pqty{ \mysymb{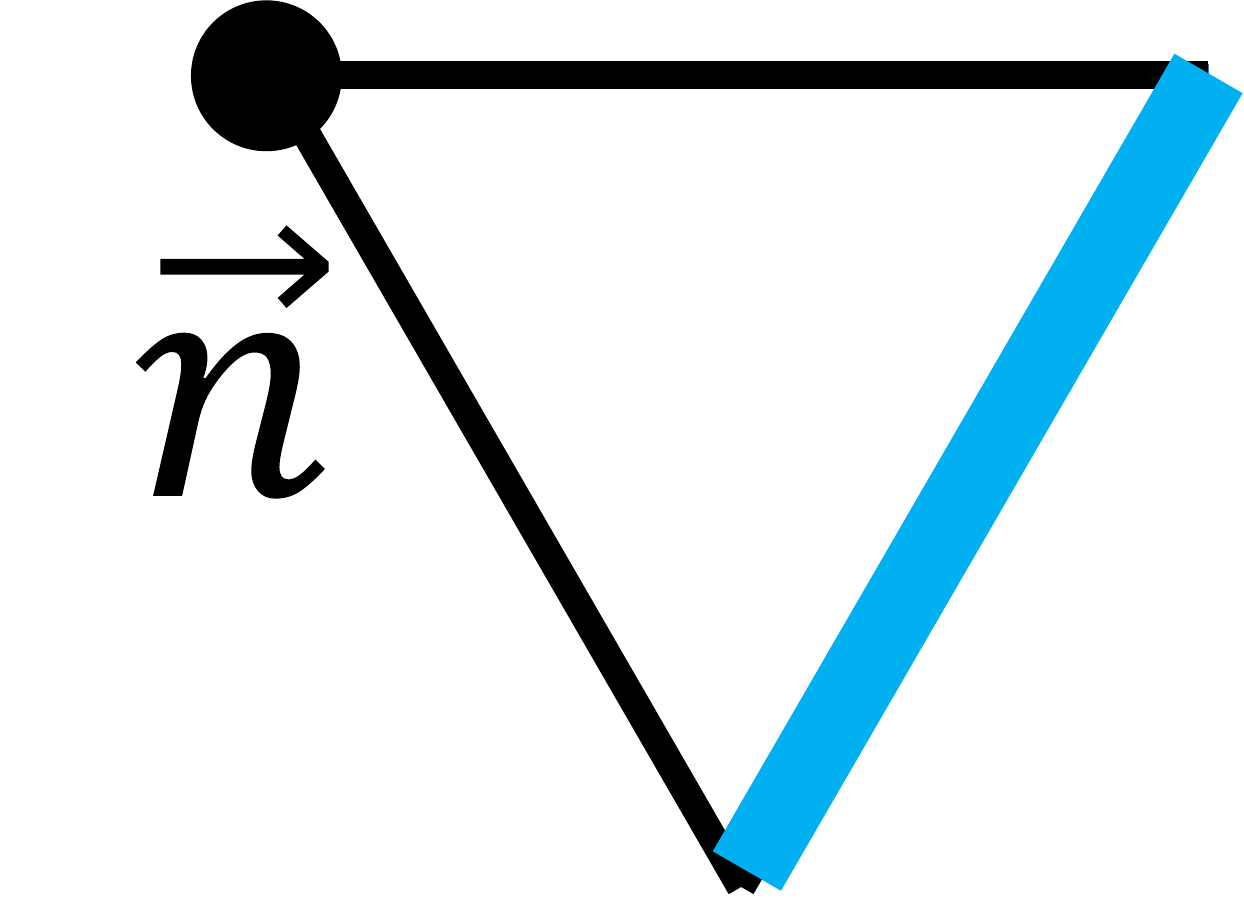}{4.0}},
    \end{aligned}
\end{equation}
\begin{equation}
    \begin{aligned}
        \label{eq:singletm2}
    \hat{\Phi}^{\dagger(2)}_1&=\sum_{\vec{n}}e^{i(-\frac{\pi}{3},\frac{\pi}{3})\cdot\vec{n}}\qty{-i\pqty{\mysymb{singlet3.pdf}{4.0}}+e^{-i\frac{\pi}{6}}\pqty{\mysymb{singlet2.pdf}{4.0}}},\\
    \hat{\Phi}^{\dagger(2)}_2&=\sum_{\vec{n}}e^{i(\frac{2\pi}{3},\frac{\pi}{3}).\vec{n}}\qty{ e^{i\frac{\pi}{6}} \pqty{ \mysymb{singlet1.pdf}{4.0}}+i\pqty{\mysymb{singlet3.pdf}{4.0}}}.\\
    \hat{\Phi}^{\dagger(2)}_3&=\sum_{\vec{n}}e^{i(-\frac{\pi}{3},-\frac{2\pi}{3}).\vec{n}}\qty{e^{-i\frac{5\pi}{6}}\pqty{ \mysymb{singlet1.pdf}{4.0}}+e^{i\frac{5\pi}{6}}\pqty{ \mysymb{singlet2.pdf}{4.0}}},
    \end{aligned}
\end{equation}
where we use the notation,
\begin{equation}
    \mysymb{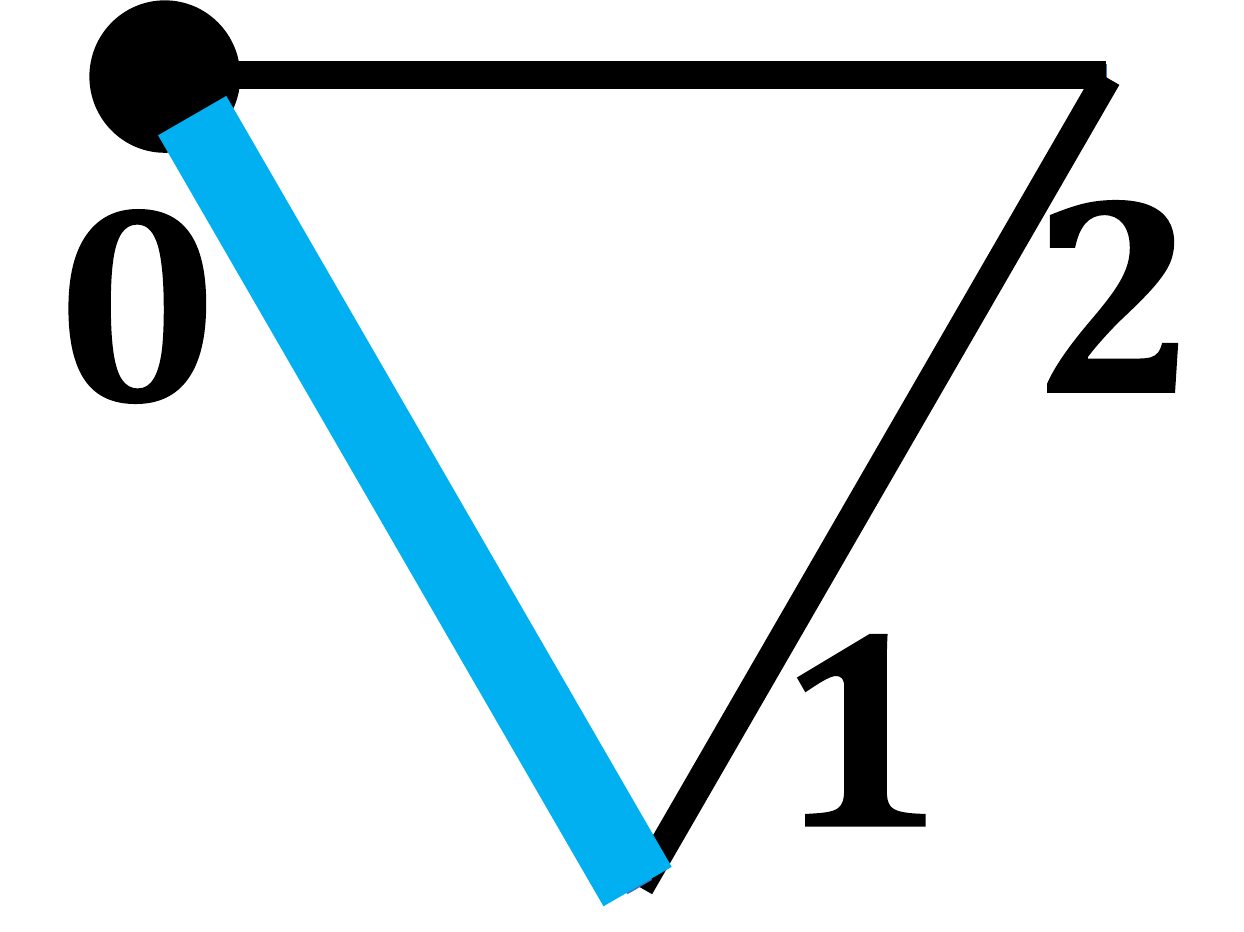}{4.0}=\dotof{0}{1}
\end{equation}
We can now use the identity Eq.~(\ref{eq:id6}) to get
\begin{equation}
\begin{aligned}
    &\comm{\tot{\hat{b}}^{(1)}}{\hat{\Phi}^{\dagger(1)}_i}=\frac{1}{2}\hat{\Phi}^{\dagger(2)}_i +\ldots,\\
    &\comm{\tot{\hat{b}}^{(1)}}{\hat{\Phi}^{\dagger(2)}_i}=\hat{\Phi}^{\dagger(1)}_i -\frac{\sqrt{3}}{2}\hat{\Phi}^{\dagger(2)}_i+\ldots,\\
    \implies &\comm{\tot{\hat{b}}^{(1)}}{v_1 \hat{\Phi}^{\dagger(1)}_i + v_2 \hat{\Phi}^{\dagger(2)}_i}=v_2 \hat{\Phi}^{\dagger(1)}_i+ \\&+ \pqty{\frac{1}{2}v_1 -\frac{\sqrt{3}}{2}v_2}\hat{\Phi}^{\dagger(2)}_i + \ldots.
    \end{aligned}
\end{equation}
If we demand proportionality already to this order, then we obtain $\mathcal{K}=v_2/v_1=0.396$. In contrast, $\mathcal{K}$ for the spin triplet monopole in Eq.~\ref{eq:tripletexp} is $\sqrt{3}$, although in theory they should be the same. The discrepancy is the result of our uncontrolled approximation to drop higher size terms, since the commutator of two high size operators can give a lower size operator (for example, Eq.~\eqref{eq:id8}). Nevertheless, using this approach we have been able to motivate why the $U(1)_{\text{top}}$ contribution to monopole momentum is $(2\pi/3,-2\pi/3)$. 

It could be a fruitful direction to assume that the coefficients $C_s$ do decay with operator size $s$ and self-consistently solve for $C_s$ using the general approach described above. Since the generators $\hat{Q}^{ab}$ are generators for emergent global \textit{internal} symmetries, na\"ively, one would expect that $\hat{Q}^{ab}$ is a sum of approximately local terms, and $C_s$ decays exponentially with size $s$. It will be interesting to verify that this is indeed the case, and if so, to determine what sets the decay length when the IR theory is conformally invariant. If this approach succeeds, it would help one to study DSLs without resorting to parton construction, and serve as a technique complementary to the one explored in~\cite{ye2021topological}.
\subsection{List of useful commutation relations}
Here, we list some useful commutation relations of various spin operators with the spin chirality, i.e. commutators of the form
\begin{equation}
    \comm{\hat{\mathcal{O}}\pqty{\{\va{S}_i\}}}{\chiof{1}{2}{3}}
\end{equation}
where $\hat{\mathcal{O}}\left[\{\vec{S}_i\}\right]$ is a local operator made of spins. 

\noindent \underline{$\hat{\mathcal{O}}$: Spin triplet made of single spin}
 \begin{equation}
 \label{eq:id1}
        \comm{\Sof{1}}{\chiof{1}{2}{3}}=i\pqty{(\dotof{1}{2})\Sof{3}-(\dotof{1}{3})\Sof{2}}
    \end{equation}
\underline{$\hat{\mathcal{O}}$: Spin triplet made of 3 spins}
\begin{equation}
    \label{eq:id8}
    \begin{aligned}
    &\comm{(\dotof{1}{2})\Sof{3}}{\chiof{1}{2}{3}}=-\frac{i}{8}(\Sof{1}-\Sof{2})+\\&+\frac{i}{4}\pqty{(\dotof{2}{3})\Sof{1}-(\dotof{1}{3})\Sof{2}}
    \end{aligned}
\end{equation}
\begin{equation}
    \label{eq:id9}
    \comm{(\dotof{1}{2})\Sof{3}}{\chiof{1}{2}{4}}=-\frac{i}{2}(\dotof{1}{4}-\dotof{2}{4})\Sof{3}
\end{equation}
\begin{equation}
    \label{eq:id10}
    \begin{aligned}
    &\comm{(\dotof{1}{2})\Sof{3}}{\chiof{1}{3}{4}}=\frac{i}{2}\left((\dotof{2}{4})\Sof{1}+\right.\\&\left.+(\dotof{3}{4})\Sof{2}-(\dotof{2}{3})\Sof{4}-(\dotof{1}{4})\Sof{2}\right)
    \end{aligned}
\end{equation}
\underline{$\hat{\mathcal{O}}$: Spin singlet made of two spins}
\begin{equation}
    \label{eq:id6}
    \comm{\dotof{1}{2}}{\chiof{1}{2}{3}}=-\frac{i}{2}(\dotof{1}{3}-\dotof{2}{3})
\end{equation}
\subsection{Remarks on notation}\label{app:notation}
We write the generator corresponding to the charge/current in square brackets. Eg: $\hat{Q}\bqty{\sigma^i \tau^j}$ and $\hat{Q}\bqty{U(1)_{\text{top}}}$.

Correspondence between the notations $\hat{Q}^{ab}$ and $\tot{\hat{Q}}\bqty{\sigma^i \tau^j}$:
\begin{equation}
    \tot{\hat{Q}}\bqty{\sigma^1}=\hat{Q}^{56}, \quad \tot{\hat{Q}}\bqty{\sigma^2}= \hat{Q}^{64}, \quad \tot{\hat{Q}}\bqty{\sigma^3}=\hat{Q}^{45}
\end{equation}
\begin{equation}
    \tot{\hat{Q}}\bqty{\tau^1}=\hat{Q}^{23},  \quad \tot{\hat{Q}}\bqty{\tau^2}=Q^{31},  \quad \tot{Q\bqty{\tau^3}}=\hat{Q}^{12}
\end{equation}
\begin{equation}
    \tot{\hat{Q}}\bqty{\sigma^i \tau^j}=\hat{Q}^{3+i,j} \text{ for } 1\leq i,j \leq 3
\end{equation}

\section{Ignoring source terms for spin singlet monopoles}
\label{app:ignore}
In this section, we argue why source terms for spin singlet monopoles potentially arising due to spatial symmetry breaking near the boundaries (see Eq.~\eqref{eq:hsource}), do not significantly affect the $U(1)_{\text{top}}$ Josephson current between two $120^\circ$ AFMs. For simplicity, let us work with the effective Hamiltonian in terms of the ordered phases alone, with the DSL integrated out, as we did in Eq.~\eqref{eq:effham1}. The source term, localized to the boundaries modifies Eq.~\eqref{eq:effham1} as follows:
\begin{equation}
    \hat{H}_{\text{new}}=\hat{H}_{\text{eff}}+ \sum_{i=1}^3\sum_{P=L,R} \pqty{V^{\text{eff}}_{i,P}\hat{\Phi}_{i,P}^\dagger+\text{ h.c.}}
\end{equation}
Note that $V^{\text{eff}}_{i,P}$ is a coupling arising under RG flow in the effective field theory due to the boundaries breaking spatial symmetries. Hence, it is small when compared to $\Gamma^{\text{eff}}_S\expval{\hat{\va{\Phi}}_{L/R}}$, which in contrast is macroscopic in the $120^\circ$ AFM. The source term leads to the following extraneous contribution to the $U(1)_{\text{top}}$ current:
\begin{equation}
\begin{aligned}
    -\pqty{\dv{\hat{b}_{\text{tot},L}}{t}}_{\text{extra}}&=i\comm{\hat{b}_{\text{tot},L}}{\sum_{i=1}^3 \pqty{V^{\text{eff}}_{i,L}\hat{\Phi}_{iL}^\dagger+\text{ h.c.}}}\\
    &=i\sum_{i=1}^3\pqty{V^{\text{eff}}_{i,L}\hat{\Phi}_{iL}^\dagger-\text{ h.c.}} 
    \end{aligned}
\end{equation}
Now, we take expectation value of the above expression. The result is proportional to the expectation value of a spin singlet monopole at the boundary of a $120^\circ$ AFM phase. The only reason this expectation value is nonzero is because of $V^{\text{eff}}_{i,L}$. Hence, $\expval{\hat{\Phi}_{iL}}$ is first order in $V^{\text{eff}}_{i,L}$. Therefore, $\expval{-\pqty{\dv{\hat{b}_{\text{tot},L}}{t}}_{\text{extra}}}$ is second order in $V^{\text{eff}}_{i,L}$, which we neglect due to the assumption that $V^{\text{eff}}_{i,L}$ is small.
\section{Formula for Raman scattering off a non-equilibrium state}
\label{app:raman}
In this appendix, we will derive Eq.~\eqref{eq:Rgeneral} for the Raman scattering rate when the spin system is not in an energy eigenstate, but in a nonequilibrium steady state. While we will have Raman scattering in mind for the sake of concreteness, our derivation applies for any scattering process. We have two systems --- light and matter. Light is used to probe matter (the DSL in our case). The full time-independent Hamiltonian is
\begin{equation}
    \hat{H}=\hat{H}_0 + \hat{V}
\end{equation}
where $\hat{H}_0$ is the Hamiltonian for the matter and light fields separately and $\hat{V}$ is the light matter coupling. Suppose that at time $t=0$, the system is in state $\ket{\psi}\otimes \ket{n_i;0}$, i.e. the matter part of the state is $\ket{\psi}\equiv \sum_l \psi_l\ket{l}$ (where $\ket{l}$ is an energy eigenstate of the matter Hamiltonian) and the light part has $n_i$ photons in a mode of frequency $\omega_i$ and $0$ photons in mode $\omega_f$. In the final state, at time $T$, the light part is in the state $\ket{n_i -1; 1}$, while the matter part is in an unknown state $\ket{f}$. The scattering rate is given by 
\begin{align}
\label{eq:Rformula}
    R&=\frac{1}{T}\sum_f \abs{\pqty{\bra{f}\otimes \bra{n_i -1;1}}\hat{U}(T)\pqty{\ket{\psi}\otimes \ket{n_i;0}}}^2 \nonumber \\ 
    &=\frac{1}{T} \sum_f \abs{\sum_l \psi_l \pqty{\bra{f}\otimes \bra{n_i -1;1}}\hat{U}(T)\pqty{\ket{l}\otimes \ket{n_i;0}} }^2
\end{align}
where $\hat{U}(T)\equiv e^{-i(\hat{H}_0 + \hat{V})T}$ is the time-evolution operator. For ease of notation, we now define
\begin{align}
    \ket{L}&\equiv \ket{l}\otimes \ket{n_i;0} \text{ and } H_0 \ket{L}=E_L \ket{L} \nonumber \\ &\text{ where } E_L\equiv E_l + n_i \omega_i\\
    \ket{F}&\equiv \ket{f}\otimes \ket{n_i-1;1} \text{ and } H_0 \ket{F}=E_F \ket{F} \nonumber \\ &\text{ where } E_F\equiv E_f + (n_i-1) \omega_i + \omega_f
\end{align}
So, Eq.~\eqref{eq:Rformula} becomes
\begin{equation}
    R=\frac{1}{T}\sum_f \abs{\sum_l \psi_l \mel{F}{\hat{U}(T)}{L}}^2.
\end{equation}
For $T>0$, 
\begin{equation}
    \hat{U}(T)= i \hat{G}_R (T) = i\int_{-\infty}^\infty \frac{d\omega}{2\pi} \hat{G}_R(\omega) e^{-i\omega T},
\end{equation}
where $\hat{G}_R(\omega)$ is the retarded Green's function for the full system (light + matter). Using the standard T-matrix formalism, we can write
\begin{equation}
    \hat{G}_R(\omega) = \hat{G}_R^{0}(\omega) + \hat{G}_R^{0}(\omega) \Tm (\omega) \hat{G}_R^{0} (\omega)
\end{equation}
where $\hat{G}_R^0 (\omega)=\frac{\mathbb{1}}{\omega^+ - \hat{H}_0}$ (here, $\omega^+\equiv \omega + i0^+$) and $\Tm (\omega)=\hat{V}+ \hat{V} \hat{G}_R^0 (\omega) \hat{V} + \hat{V} \hat{G}_R^0 \hat{V} \hat{G}_R^0 \hat{V} + \ldots$.

Clearly, $\hat{G}_R^0$ cannot induce a transition that changes the number of photons; only the second term involving $\Tm$ can do so. Thus, we get the following scattering amplitude
\begin{equation}
    \mel{F}{\hat{U}(T)}{L}=i \int_{-\infty}^{\infty} \frac{d \omega}{2\pi} e^{-i \omega T} \frac{\mel{F}{\Tm (\omega)}{L}}{\pqty{\omega^+ - E_F}\pqty{\omega^+ -E_L}}
\end{equation}
The $\omega$ integral should be closed in the lower half plane for convergence. This integral will pick up poles at $E_F - i0^+$ and $E_L - i0^+$. The poles of $\Tm (\omega)$ will not play a role under the assumption $T \gg 1/(E_F- E_M)$, which is the regime of interest since we wish to consider the large $T$ limit. Here, $E_M$ is the total energy (light + matter) of any level $M$ such that $\mel{F}{\hat{V}}{M}\neq 0$. In such a large $T$ limit, one can expand out $\Tm (\omega)$ and see that our assumption is justified. So, we get
\begin{align}
    \mel{F}{\hat{U}(T)}{L}=&-2i e^{-i(E_F + E_L)T/2} \ \frac{\sin ((E_F - E_L)T/2)}{E_F - E_L}\nonumber\\ &\times \mel{F}{\Tm (\omega=E_F)}{L}\\
    \label{eq:ampformula}
    \approx & -2\pi i e^{-i E_F T} \delta(E_f + \omega_f - E_l -\omega_i)\nonumber \\ &\times \mel{F}{\Tm (\omega=E_F)}{L}
\end{align}
Now, $\mel{F}{\Tm (\omega=E_F)}{L}$ is the same operator that appears in the equilibrium calculation in~\cite{shastry1990theory,ko2010raman}. As shown there, up to a constant of proportionality 
\begin{equation}
    \mel{F}{\Tm (\omega=E_F)}{L} = \mel{f}{\hat{M}}{l}
\end{equation}
i.e. the above matrix element for the full system is proportional to a matrix element of the matter part alone. $\hat{M}$ has been calculated in~\cite{shastry1990theory,ko2010raman} and depends on the initial and final polarizations of light, momentum transferred by light and the lattice of the matter system. We have presented the leading order expression for $\hat{M}$ in Eq.~\eqref{eq:fleuryloudon}. Substituting Eq.~\eqref{eq:ampformula} into Eq.~\eqref{eq:Rformula}, we get
\begin{align}
    R\approx&\frac{1}{T} \sum_f \sum_{l,l'} \psi_{l'}^* \psi_l (4\pi)^2 \delta (E_f + \omega_f -E_l - \omega_i)\nonumber\\ &\times\delta(E_f+\omega_f-E_{l'}-\omega_i)\mel{l'}{\hat{M}^\dagger}{f}\mel{f}{\hat{M}}{l} \nonumber\\
    =&\frac{1}{T} \sum_f \sum_{l,l'} \psi_{l'}^* \psi_l (4\pi)^2 \delta (E_f + \omega_f -E_l - \omega_i)\nonumber\\ &\times \delta(E_{l'}-E_l)\mel{l'}{\hat{M}^\dagger}{f}\mel{f}{\hat{M}}{l}\\
    =&\lim_{T\to \infty} \frac{1}{T} \sum_f\int_{-\frac{T}{2}}^{\frac{T}{2}}d t_0  e^{i(E_{l'}-E_l)t_0}\nonumber \\&\times \int_{-\infty}^{\infty}dt  e^{i(E_f+\omega_f - E_l - \omega_i)t}\nonumber \\ &\times \sum_{l,l'} \psi_{l'}^* \psi_l  \mel{l'}{\hat{M}^\dagger}{f}\mel{f}{\hat{M}}{l}
\end{align}
where in the last equation, we used the Fourier representation of the $\delta$ function. Now, we can associate the phases in the above equation with the phases coming from time evolution to simplify it as follows:
\begin{align}
    R=& \lim_{T\to \infty}\frac{1}{T}\sum_f \sum_{l,l'}\int_{-T/2}^{T/2} dt_0 \int_{-\infty}^{\infty} dt e^{i(\omega_f - \omega_i)t} \nonumber \\ &\times \psi_{l'}^* \psi_l \mel{l'}{\hat{M}^\dagger e^{i\hat{H}_0 t_0}}{f}\mel{f}{e^{i\hat{H}_0 t}M e^{-i\hat{H}_0 (t+t_0)}}{l} \nonumber\\
    =&\lim_{T\to \infty} \frac{1}{T}\int_{-T/2}^{T/2} dt_0 \int_{-\infty}^{\infty} dt e^{i(\omega_f - \omega_i)t}\nonumber \\ &\times \mel{\psi}{\hat{M}^\dagger (t_0) \hat{M}(t+t_0)}{\psi}
\end{align}
where $\hat{M}(t)=e^{i\hat{H_0}t}\hat{M}e^{-i\hat{H}_0 t}$. This completes the derivation of Eq.~\eqref{eq:Rgeneral}.

\bibliography{MonopoleRef}

\end{document}